\documentstyle[preprint,aps]{revtex}
\usepackage{psfig}

\begin{document}

\title{Design and Folding of Dimeric Proteins}
\author{Guido Tiana$^{1,2}$ and Ricardo A. Broglia$^{1,2,3}$}
\address{$^1$Department of Physics, University of Milano,}
\address{via Celoria 16, 20133 Milano, Italy,}
\address{$^2$INFN, Sez. di Milano, Milano, Italy,}
\address{$^3$The Niels Bohr Institute, University of Copenhagen,}
\address{Bledgamsvej 17, 2100 Copenhagen, Denmark}
\date{\today}
\maketitle

\begin{abstract}
In a similar way in which the folding of single--domain proteins provide an important
test in the study of self--organization, the folding of homodimers 
constitute a basic challenge in the quest for the mechanisms which are at the
basis of biological recognition. Dimerization is studied by following the evolution of two
identical 20--letter amino acid chains within the framework of a lattice model and using
Monte Carlo simulations. It is found that when design (evolution pressure) selects few,
strongly interacting (conserved) amino acids to control the process, a three--state folding
scenario follows, where the monomers first fold forming the halves of the eventual dimeric
interface independently of each other, and then dimerize ("lock and key" kind of association).
On the other hand, if design distributes the control of the folding process on a large number
of (conserved) amino acids, a two--state folding scenario ensues, where dimerization takes
place at the beginning of the proces, resulting in an "induced type" of association.
Making use of conservation patterns of families of analogous dimers, it is possible to compare
the model predictions with the behaviour of real proteins. It is found that theory provides
an overall account of the experimental findings.
\end{abstract}

\section{Introduction}

A large number of proteins perform their biological activity under the shape of dimers (or
oligomers). A dimer is a protein whose native conformation 
is a globule build out of two disjoint chains.
Depending whether the two chains have the same sequence or not, they are referred to as
homodimers or heterodimers. Notwithstdanding this difference, they have been observed
to fold through two major paradigms\footnote{A third paradigm, that of "domain--swapping", 
has been described in detail by Eisenberg and coworkers \protect\cite{eisenberg}.
In the present work we concentrate only on the other two kinds of dimerization.} \cite{tsai,sh_nsb,mateu}. Some of the known dimers fold
according to a three--state mechanism (D$\rightarrow$I$\rightarrow$N), 
where first the denaturated chains of the monomers (D) assume 
conformations rich of native structures independently of each
other (I: folding intermediate), and subsequently the two parts come together to form the dimer (N: native). 
This is the case, for example, of aspartate aminotransfease \cite{asp_d}, where one can observe three populated
species, namely unfolded monomers (D), partially folded monomers (I) and folded dimers (N). A different behaviour is
displayed by, for example, P22 arc repressor \cite{arc_d}, whose chains dimerize without populating any monomeric native--like
intermediate (two--state process, D$\rightarrow$N). 
In this case one can only identify the unfolded monomers (D) and the native dimers (N).

The aim of the present work is to achieve, using model calculations, 
a basic understanding of the folding mechanism of dimers
based solely on energetic arguments, as was already done in the case of small, single domain proteins (monomers).
In this case, it was found that good folders are those sequences whose total energy
in the ground state conformation lies below a threshold value $E_c$ \cite{sh_nature,prl}. 
This threshold energy is solely determined
by the number of amino acids forming the chain, the standard deviation of the contact energies used 
in designing the sequences and by their composition.
It is equal to the lowest energy that random heteropolymers of the same length and composition can achieve
when they compact into the ensemble of conformations
structurally dissimilar\footnote{That is, the structures having a small value ($\leq 0.4$) of the
similarity parameter (order
parameter) $q$, defined as the ratio between the number of native contacts present in the
structure to which the sequence has compacted, and the total number of contacts of the
native structure used to design the sequence. In the case of dimers, it is convenient to introduce
two similarity (order) parameters for any given conformation, namely $q_1$ and $q_t$ which
correspond to the relative value of native contacts within each chain and across the two chains,
respectively.} to the native conformation.
 
Aside from being responsible for the thermodynamical uniqueness and stability of the ground state, 
the low--energy character of single domain good folders is also essential for their dynamics. The
only way a small, single domain lattice designed protein can display an energy below $E_c$ 
is by positioning (few) strongly interacting amino acids
in some key sites of the protein \cite{hierarchy,jchemphys1,nik1,nik2}. These sites,
called "hot" sites in ref. \cite{jchemphys1} participate in the
formation of local elementary structures (LES), which bias the chain to its native conformation \cite{modellino,aggreg}
and which build the (post--critical) folding nucleus when they get together, responsible not only for the stability
of the protein\cite{sh_nucleus}, but also of its fast folding ability. 
Amino acids in these sites are highly conserved in evolution
\cite{leo} and determine the topology of the space of folding sequences \cite{isles}.

In the following we will characterize the dynamic behaviour of model homodimers with respect
to the energetic properties of their ground state conformation.
Since a reliable potential function for residue--residue interaction is
not available, we can compare the results of model calculations to real proteins only through
the analysis of conservation patterns in families of analogous\footnote{That is, sequences folding
to the same dimeric conformation but not connected by evolution. Couples of sequences whose similarity
is lower than $20\%$, measured as percentage of matching residues, are likely to
be analogous \protect\cite{sanders}.} dimers. Such an analysis has been performed for
the analogs of aspartate aminotransfease and of P22
arc repressor, representative examples of a three--state and a two--state folding homodimer, respectively. 
The corresponding results are displayed in Figs. \ref{arc} and \ref{asp}, 
and discussed in Section V, where they are compared with the conservation patterns found in lattice model proteins. 
 
\section{Design of Sequences}

The model we use to study homodimers has been largely employed with success  \cite{sh_nature,prl,sh_nucleus,go,dill} in the study of monomers. This is 
because, in spite of the strong simplifications introduced in the description of the proteins, aimed at making
feasible dynamical simulations of the folding process,
the model still contains the two main ingredients which are at the basis of the
distinctive properties of proteins: polymeric structure and disordered interactions 
among the amino acids\cite{wolynes}. The model is used to study the general thermodynamical and kinetical properties of notional dimers, independently on details concerning secondary structures, side chains, etc. Of course, all these details may prove of relevance when addressing specific questions, as was e.g. done in refs. \cite{mohanty,vieth} concerning the oligomerization equilibrium properties of leucine zippers.

According to the model, a protein is a chain of beads on a cubic lattice, each bead representing
an amino acid (selected from twenty different types)
which interacts with its nearest neighbours through a contact potential
with numbers taken from the statistical analisys of real proteins carried out by Miyazawa and Jernigan
in ref. \cite{mj} (for details about the model see, e.g., refs. \cite{jchemphys1,sh_nucleus}).

In the case of single domain proteins it is, in principle, simple to design sequences which fold
to a given target conformation.
Due to the fact that the thermodynamical and kinetical behaviour of a protein are essentially
determined by its total energy, it is possible to design folding sequences by searching in the
space of sequences for those having energy lower than $E_c$.
Using a Monte Carlo algorithm, a sequence with energy $E$ in the target conformation 
has a probability to be selected proportional to $\exp(-E/\tau)$,
where $\tau$ is an intensive variable which plays the role of temperature and
gives the degree of bias towards low--energy sequences.
In the evolutionary context, $\tau$ has the meaning of selective pressure with respect to
the protein ability to fold \cite{isles}: the lower is the value of $\tau$, the stronger is
this pressure and the better are
the folding properties of the selected sequences. In particular, for values of $\tau$ lower than
the temperature $\tau^c\equiv (\partial S/\partial E|_{E=E_c})^{-1}$ (which is the
temperature at which the mean energy is exactly $E_c$), the average energy of
the selected sequences is lower than $E_c$. Consequently one obtains in this way sequences with a unique
and stable native conformation and able to find it rapidly. In other words, the designed sequences display a unique
ground state (native conformation) with energy $E_N<E_c$ into which they fold fast \cite{hierarchy,sh_des}.

In order to design homodimers, we first set a target conformation built out of two identical parts
(chosen equal to the native structure of a 36mer widely used in the literature in studies
of small, single domain proteins \cite{jchemphys1,sh_nucleus}),
having a face in contact (see Fig. \ref{native}).
In the present model the monomers have been chosen to be mirror images of each other, consequently
the overall structure is symmetrical with respect to the interface, a feature which simplifies 
the computational handling of the protein\footnote{ 
Strictly speaking, real homodimers 
are composed of two identical (not symmetrical) monomers.
On the other hand, from an energetic point of view our choice is of no consequence.
This is because the energy content of a model conformation is solely characterized by its interaction map. Concerning the entropy of the dimer, our choice of native conformation implies a (two--fold) degenerate ground state. Because, this doubling of the degeneracy affects equally all the states of the system, it is of no consequence for the results reported below.}.

The second step in the homodimer design is to choose 
a realistic ratio between the different kinds of amino acids\cite{creighton} and two evolutionary
temperatures, $\tau_1$ and $\tau_t$, which control the amino acids in the bulk 
and at the interface, respectively. We then use a multicanonical sampling algorithm (see Appendix A)
to select a set of sequences $\{s\}$, according to the distribution of probability
\begin{equation}
p(\{s\})=\frac{1}{Z}\exp\left(-\frac{H_1(\{s\})}{\tau_1}-\frac{H_t(\{s\})}{\tau_t}\right),
\end{equation}
where $H_1(\{s\})$ and  $H_t(\{s\})$ are the energies associated with the contacts between residues belonging
to the same monomer and between residues across the interface, respectively, and $Z=\sum_{\{s\}}
\exp(-H_1/\tau_1-H_t/\tau_t)$ is the partition function. 
During the sampling process, couples of residues belonging to one of the two monomers
are swapped, the same swap being repeated on the other monomer, in such a way that the sequence
remains identical in the two parts of the dimer. Moreover, in this way the overall concentration of
the different kinds of residues is kept constant.

Starting from a random sequence displaying a realistic composition (i.e. the average ratio
of different kinds of amino acids as found in natural proteins \cite{creighton}), we make 
$10^6$ swappings, sufficient to ensure a stationary distribution of energy, before selecting
a sequence, repeating this process $10^4$ times, to obtain a statistically representative set
of designed sequences $\{s\}$. Consequently, in constructing this set we sample only $10^{10}$
sequences of the total of $20^{36}=10^{46}$ sequences. In keeping with the fact that each of 
the $10^4$ selected sequences are separated from each other by $10^6$ swappings, it is 
reasonable to assume that $\{s\}$ is a set of evolutionary uncorrelated sequences whose properties
(e.g., conservation patterns) can be compared to those of non--homologous sequences (analogous) families
of real proteins.

The use of two different temperatures $\tau_1$ and $\tau_t$ to select residues in the bulk and at the interface
respectively, corresponding to two different evolutive pressures, allows one to control how the energy,
and the amino acid conservation pattern,
is distributed inside the dimer. Details and caveats concerning the details and limitations
of the design procedure are given
in Appendix B. Examples of sequences selected at different evolutionary temperatures to fold to the
structure displayed in Fig. \ref{native} are listed in Table I.

\section{Equilibrium Properties of Sequences}

In the present and in the next Sections we shall study the properties
and the behaviour of the designed sequences in conformation space,
making use of long Monte Carlo runs. In what follows we shall (mainly)
concentrate on the thermodynamical properties of these sequences, while
in Sect. IV we dwell primarily on their dynamical behaviour.

We first proceed to the calculation of the ground state energy of the designed sequences.
Because a complete enumeration of all possible homodimer conformations is out of question,
the thermodynamical properties of the sequences selected at different 
values of $\tau_1$ and $\tau_t$ (see e.g. Table I)
are analyzed through a standard Metropolis algorithm \cite{metropolis} 
in the conformational space at fixed temperature $T$ 
(temperature in "real" conformational
space, not to be confused with the "evolutionary" temperatures 
$\tau_1$ and $\tau_t$ in the space of sequences).
The simulations were performed with periodic boundary conditions, i.e. in a 
cubic Wigner cell of dimension $L$ (cf. App. C). The need for using a finite volume
arises from the fact that, in an infinite volume each conformation with
disjoint chains have an infinitely negative free energy, due to its
infinite translational entropy\footnote{In fact, the entropy of the disjoint system
is proportional to $\ln\Omega^2$, while that of the dimer is proportional
to $\ln\Omega$, $\Omega$ being the volume allowed to the protein to move in.}.

Long Monte Carlo simulations ($5\cdot 10^8$ MC steps) were performed for each pair of (identical) designed chains. Two outcomes were observed: (A) folding into the dimeric native state, (B) aggregation into a set of disordered clumps.
In what follows we shall only analyze the outcome (A), while outcome (B) 
will be discussed in a forthcoming paper \cite{dimernew}.
In Fig. \ref{dyn_phases} we show the results obtained by carrying out the MC
simulation at $T=0.24$, for which the fractional population of the native
state is $55\%$. Full simbols correspond to sequences displaying the behaviour (A),
while empty symbols correspond to sequences displaying the behaviour (B).
The solid line  indicates the loci where the values of $\tau_1$ and $\tau_t$ correspond to the energy $E_c=-35$.
We note that this line delimits the area in which the corresponding designed sequences 
display the behaviour of type (A).
This is not surprising, considering the fact that the lowest energy of
a random conformation with the same number $n=92$ of contacts as the target
conformation
is $E_c=-n\sigma (2\ln\gamma)^{1/2}=-34.7$ (evaluated in the approximation of 
the random energy model \cite{derrida}, $\sigma=0.3$ being the standard deviation
of the interaction matrix elements and $\gamma=2.2$ the effective coordination
number \cite{orland}). As in the case of single domain proteins, conformations
with energy lower than $E_c$ have not to compete with the sea of random
structures, and consequently are good native conformations (unique and stable)\footnote{From 
the experience accumulated on lattice designed, small single--domain
proteins, it was learned that the condition that good folders correspond to sequences displaying a 
sufficiently low energy $E_{design}$ in the native conformation (i.e. $\xi=(E_{design}-E_c)/\sigma\gg 1$,
quantity closely connected with the z--score \cite{bowie}, and where
$\sigma$ is the standard deviation of the interaction matrix) is actually more general \cite{hierarchy}.
In fact, because these monomers fold through local elementary structures (LES) stabilized by 
few (hot), strongly interacting amino acids, and because these amino acids are conserved for all
sequences with $E_{design}<E_c$, one can restate the large energy gap paradigm of good folders as follows:
sequences which display a very small number of hot amino acids
and which conserve, in any way, the energy gap \protect\cite{prl}.} .

In what follows we shall discuss, making again use of the results of long Monte Carlo runs\footnote{The
runs were carried out in a cubic Wigner cell of sidelength $L=7$ and periodic boundary conditions
because this was found the most appropriate from numerical considerations (cf. App. C).}, 
the properties of two of the sequences shown in Table I, namely
sequences $\#1$ and $\#3$, chosen as representative examples of chains building homodimers which
fold according to a two-- and to a three-state mechanism, respectively.
They were both designed at low $\tau_t$ and $\tau_1$, in the first case
with $\tau_t<\tau_1$ and in the second case with $\tau_t>\tau_1$.
In both cases, the designed sequences display a first--order transition in conformational space
from the denaturated state into the
native dimeric structure, as in the case of single domain proteins, as testified by the discontinuous
behaviour of the similarity (order) parameters at the critical temperature $T^c$.
In other words, the order parameters $q$ display at $T=T^c$ a double peak behaviour.

As a rule, the critical temperatures $(T^c)_1$ and $(T^c)_t$
in conformational space will be different. In particular, 
in the case of sequence $\#1$ of Table I, $(T^c)_1<(T^c)_t$, in keeping with the fact that this sequence
was designed setting $\tau_1$ larger than $\tau_t$. This property has the consequence that 
the native state is stable at 
temperatures $T<(T^c)_1\approx 0.24$, $(T^c)_1$ being the critical temperature,
at which the area of the peak associated to the unfolded
state ($q_1$ small) is equal to that associated to the native state ($q_1=1$), and that 
dimerization happens already at 
$T<(T^c)_t\approx 0.26$. The fact that $(T^c)_1<(T^c)_t$ indicates that
while it is possible to find a situation (at $T$ such that $(T^c)_1\leq T<(T^c)_t$)
in which the two chains build the native interface but the monomers are not 
folded (cf. e.g. Fig. 5(b) ), there is not
a temperature at which the monomers are folded and separated (cf. Fig. 5(a) ).

Also sequence $\#3$ undergoes a first order transition, but in this case the critical
temperature $(T^c)_1$ associated with the bulk is larger than that associated with
the interface ($(T^c)_1=0.25>(T^c)_t=0.22$). This indicates that for sequences selected at
comparatively high selective temperature $\tau_t$ ($=10\tau_1$, cf. Table I), 
there exists a phase where the two chains are folded
but separated (cf. Fig. 6).

Another interesting thermodynamical property of a dimer is the localization
of the sites which are mostly responsible for the stabilization of the
native state. These sites can be identified by the change in energy $\Delta E_{loc}$ 
that  mutations induce in the native conformation \cite{jchemphys1}. 
For a given site $i$, 19 mutations can be carried out. Because $\Delta E_{loc}(i)$ may
change markedly depending on the point mutation introduced in the protein,
it is useful to define the average value $\overline{\Delta E_{loc}(i)}$ taken over
all possible substitutions. In keeping with the result of studies of single
domain protein \cite{jchemphys1}, in particular of the sequence $\#6$ of Table I, "hot" sites
(sites which are most sensitive to point mutations, which as a rule, if mutated, 
denaturate the protein) are those sites for which $\overline{\Delta E_{loc}(i)}>\mu+2\sigma$,
where $\mu$ is the average value of $\overline{\Delta E_{loc}(i)}$ ($i=1,...,N$) taken over all the sites of the chain, while $\sigma$
is the associated standard deviation. "Cold" and "warm" sites in the nomenclature of
ref. \cite{jchemphys1} are those sites for which  $0\leq\overline{\Delta E_{loc}(i)}\leq\mu$ and
$\mu\leq\overline{\Delta E_{loc}(i)}\leq\mu+2\sigma$ respectively\footnote{Note that this general definition essentially coincides with that given in ref. [10] for the case of S$_{36}$, i.e. $\overline{\Delta E_{loc}(i)}<1$ (cold),
 $1\leq\overline{\Delta E_{loc}(i)}<2$ (warm), $1\overline{\Delta E_{loc}(i)}>2$ (hot), in keeping with the fact that, in this case, $\mu\approx 1$ and $\sigma\approx 0.7$.}.

In Fig. 7 we display the values of $\overline{\Delta E_{loc}(i)}$ associated with
sequence $\#1$, $\#3$, $\#5$ and $\#6$ (S$_{36}$ monomer) of Table I. A marked difference
in the pattern distribution of hot and warm sites associated with sequence  $\#1$ (two--state
folding) and $\#3$ (three-state folding) is observed. 
In fact, sequence $\#1$ (Fig. 7(a) ) displays no hot sites and double as many warm sites than sequence $\#3$ (Fig. 7(b) ), the properties of the sites of this last chain  in the native conformation being essentially
those found in the study of the isolated monomer S$_{36}$ (Fig. 7(d), cf. also \cite{jchemphys1}). Furthermore,
the amino acids occupying the warm sites of chain $\#1$ are, in average, more strongly interacting (2.4) than those associated with the warm sites of chain $\#3$ (1.9), althoug still much less than hot sites of this chain (3.1).
In other words, in sequence $\#3$ most of the binding energy is concentrated in 
few "hot" sites, while in the case 
of sequence $\#1$ the stabilization energy is spread more homogeneously throughout
the monomers. 

\section{Dynamics of dimerization}

We now concentrate our attention, making again use of the results of the Monte
Carlo simulations already discussed in the last Section, on the dynamics of the
process that leads the system from a random conformation to the native dimer
(for details and
caveats see Appendix B). The first issue to assess is whether the dimeric
native state is accessible on a short time scale (short with respect to the
random search time, $\sim 10^{50}$ MC steps). For each of the sequences selected
at various evolutionary temperatures (some of them being listed in Table I), 
20 simulations of $60\cdot 10^6$ MC steps
have been performed, recording the first passage time, i.e. the folding time
$t_f$ (cf. Table II) into the dimeric native
conformation. Each set of simulations has been repeated at three temperatures,
$T=0.24$, $T=0.28$ and $T=0.32$.

We shall first discuss the case of sequences optimized 
at very low values of
both $\tau_1$ and $\tau_t$ and with $\tau_t<\tau_1$, so that the interface energy is
close to its global minimum (solid squares in Fig. \ref{dyn_phases}).
This design procedure was followed in the expectation to obtain sequences which fold through a two--state
process, namely first dimerization and then folding. To check this scenario, we have
determined through dynamical MC simulations the distribution of $q_1$ (and $q_t$) 
associated with sequence of type $\#1$ (cf. Table I), and calculated
at the instant in which $q_t$ (or $q_1$) reach, for the first time, a value of
the order of one\footnote{For
computational easiness, and because the parameters $q_1$ and $q_t$ vary discontinuously as a function
of the number of MC steps, we have used the MC step at which a value of $q_1$ or $q_t$ larger
than $0.7$ is reached for the first time to define the instant at which to calculate the value of the other
order parameter (and thus the associated distribution) $P(q_t)$ or
$P(q_1)$ respectively.}. The results displayed in Fig. 8(a) show 
that, by the time the bulk of the two chains fold, $q_t$
is essentially equal to 1, indicating that the interface between the two chains is already in place. On the
other hand, at the time in which $q_t$ acquires for the first time a value  $\approx 1$, $q_1$ displays a rather flat
distribution, its average value being $0.4$.
One can conclude that sequence $\# 1$ of Table I first dimerizes
and then folds.

The fact that in the present case the folding time
$\tau_f$ increases with temperature (cf. Table II) indicates that the dimerization process is not diffusion limited. 
That is, $\tau_f$ is determined not only by the time
needed by the two interfaces to come in contact, but also by the stability of 
the interface structure.
In first
approximation, the diffusion time $\tau_D$ for the interfaces to meet, in the approximation that two interfaces search for
each other randomly in the space of configurations, is $\tau_D\approx (6\cdot 4\cdot V)=8200$ steps,
where 6 is the number of faces of an ideal cubic conformation, 4 takes into account the rotational
symmetries of the faces and $V=7^3$ is the volume of the system.
This time is much shorter than the folding time. 
On the other hand,
at $T=0.28$ the relative population of an isolated fragment containing the 12
interface residues in place (forming the surface of one of the two monomers)
has been calculated to be $p_{int}=0.018$ (determined through a MC simulation), 
so that the probability to have the two
interfaces structured at the same time is $p_{int}^2=3.24\cdot10^{-4}$. Consequently, 
in this simple model, the folding
time is predicted to be
$t_f=t_D\cdot p_{int}^{-2}\approx 2.5\cdot 10^{7}$, which agrees well with the value found in 
detailed MC simulations and displayed in Table II.
This picture explains also the increase of aggregation probability with temperature. In fact,
the higher is the temperature, the longer is the time that partially structured surfaces which
eventually dimerize move
around in configurational space and can bind to wrong partners, causing aggregation. 

Also the behaviour of unfolding times agrees with the picture of folding--after--dimerization.
In fact, starting from the native conformation, the average decay of $q_t$
with respect to time is (cf. Fig. \ref{unfold})  $t_{t}=5.83\cdot 10^5$ at $T=0.28$
(the values for other temperatures are listed in Table II). On
the contrary, the distribution of $q_1$
is best fitted by a stretched exponential in the form
$q_1(t)=\exp(-t^\beta/a)$, with $\beta=0.52$ and $a=1470$,
the average time being $t_1=7.99\cdot
10^5$. The fact that $q_1(t)$ does not follow an exponential
law indicates that the unfolding of each
of the two chains is not a process which depends only
on its internal (intra monomer) contacts, like e.g. in the case
of single domain proteins, cf. \cite{jchemphys1}, but is
subordinated to an external event. In keeping with this fact, and because
the detaching time $t_t$ 
(the characteristic decay time of $q_t$) is shorter than $t_1$, one can conclude that the
breaking of the native bonds associated with the internal structure of the two chains 
is a consequence of their detaching or, in other words, that the stability of
the monomer structures relies on the presence of the interface.

We now turn our attention to the sequences selected at $\tau_1<\tau_t$ in
order to be close to the global minimum of the bulk energy ($E<-16$) and 
corresponding to the solid triangles in Fig. \ref{dyn_phases}. Sequence $\# 3$ of Table I 
is an example of this class of designed homodimers.
The dynamical behaviour of this sequence
is quite different from that of sequence $\#1$, although at $T=0.28$ it folds with probability 
$0.4$ to its native conformation, in an
average time (first passage time (FPT) ) of $t_f=3.2\cdot 10^7$ (cf. Table III) MC steps, quantities which is quite
similar to those associated with sequence $\# 1$ ($p_N=0.45$ and $t_f=2.4\cdot 10^7$ MC steps, respectively). 
The analysis of the time dependence 
of $q_1(t)$ and $q_t(t)$ shows (cf. Fig. 8(b) )
that for sequence $\#3$ $q_t$ is very small at the time in which $q_1\geq 0.7$, while
at the time at which $q_t\geq 0.7$, $q_1$ is already essentially equal to 1, indicating
a three--state scenario and the esistence of an intermediate state where the two
monomers are folded and not dimerized.
That is, sequence $\# 3$ first fold to its 
monomeric native conformation and then forms the dimer. 

In the folding process, this sequence follows 
the hierarchical path typical of monomeric model proteins \cite{modellino,hierarchy}, 
building first local elementary structures (LES), then the (postcritical) folding nucleus,  finding 
the monomeric native conformation shortly after. In addition, in the present case, there is a further step 
which consists in the association of the two monomers to build the dimer (to be noted that 
sequence $\# 3$ has the same local elementary structures and folding core than the monomeric 
protein studied in Ref. \cite{modellino}, as testified by the fact that the hot 
sites are essentially the same (cf. Figs. 7(b) and (d) )).

The same sequence, when studied as an isolated chain at the same temperature, folds in an average time
 of $0.9\cdot 10^6$ steps, indicating that the time limiting step is the association (it takes 
 $2.4\cdot 10^7$ steps). As temperature increases, the folding time decreases slightly (see Table III). 
This decrease is much less pronounced than the decrease in the diffusion time as measured by the inverse of the diffusion coefficient $D^{-1}$ (which is exponential with
 temperature, see third column in Table III), indicating that the dimerization time is not purely
 diffusion limited. On the contrary, the stability of the interface (which depends exponentially
 on temperature as well) plays an important role in the dimerization time.

To check the generality of the results discussed above, we have repeated the design and the folding for another
three dimensional conformation, where the native structure of each of the 
monomers is the same, but the interface involves
the opposite face than in the case of the dimeric conformation shown in Fig. 3. 
While the folding time of sequences which first fold and then aggregate is
similar to that of sequence $\#3$, the folding time for sequences behaving according to
a two--state mechanism is substantially higher (of the order of $10^8$
MC steps). This is due to the much lower stability of the interface
built out of the two, rather unspecific surfaces stabilized by the contacts
between residues lying rather distant from each other along the chain (e.g. 19--32, 17--32),
while in the previous case the surfaces were pieces of incipient $\beta$--sheets stabilized by
quartets of residues (e.g. 3--6, 7--10).
Nonetheless, also in this case it is possible to
find sequences which fold according to the two paradigms discussed above.

Summing up, it is found that if the folding of the homodimer is controlled by stable 
LES (cf. ref. \cite{hierarchy})
stabilized by few hot, strongly interacting amino acids, the folding time decreases with
temperatures, as the folding of the protein is (post--critical) folding--nucleus--formation controlled.
In fact, the folding nucleus is the result of the docking of the LES in the appropriate way, and thus
controlled by the diffusion time of the LES, a process which is speed up by increasing the temperature.
This happens until one reaches a temperature at which the stability of LES is affected, 
beyond which the folding time rises again. Since the stabilization energy of LES is much larger
than the interaction energy between pairs of amino acids \cite{modellino}, there exists a large
window of temperatures at which folding time is short. 
This scenario reflects, to a large
extent, the temperature dependence of the folding time of monomers of which the homodimer is made of (cf. Fig. 10).

The situation is quite different in the case of a dimer built out of two sequences of
type $\#1$, which first dimerize and then fold (two--state process). In this case, the folding of the
system is not controlled by LES. This
is because, were one to put "hot" amino acids (needed to stabilize the LES) 
on the surfaces which dimerize, these surfaces will be
so reactive that the system will aggregate with high probability (open circles in the low--$\tau_t$ region
of Fig. 4). But if the dimerizing surfaces do not contain "hot" amino acids, the folding of the remaining amino acids ("volume")
cannot depend on LES either, otherwise the system will behave more like sequences of type $\#3$.
Consequently, if both the dimerizing surfaces and the remaining part of the ("volume") system are
marginally stable, the binding energy being essentially uniformly distributed among all the pairs
of nearest neighbours amino acids (low $\varphi$--values associated essentially with all sites), 
the fast folding temperature window is expected to be much narrower than in
the case of homodimer based on sequence of type $\#3$. In fact, already at low temperature ($T=0.24$),
the system based on sequences of type $\#1$ is found to be passed the minimum of the curve shown in
Fig. 10, and one observes an increase of the folding time as a function of $T$ (cf. Table II).

\section{Results for real proteins}

A possible connection between the model of homodimer folding discussed above
and real proteins can be achieved by studying the degree of amino acid conservation 
in each site of analogous dimers.
This can be done by calculating the entropy in the space of sequences 
in each site, given by
\begin{equation}
S(i)=-\sum_{\sigma=1}^{20} p_i(\sigma)\log p_i(\sigma),
\end{equation}
where $p_i(\sigma)$ is the probability of finding the amino acid of kind $\sigma$ at site $i$, 
a probability to be calculated over a large number of sequences folding to the same conformation.

Within the present model, the calculation of $S(i)$ is straightforward. 
Given a dimeric native conformation (Fig. \ref{native}), 
and a couple of evolutionary temperatures $\tau_1$ and $\tau_t$, one selects from the
corresponding set of designed sequences a representative sample (e.g. $10^4$) 
of them. Since these sequences are aligned, it is possible to calculate the associated values of $p_i(\sigma)$ 
and, consequently, $S(i)$. The entropy at each site ranges from $0$, if only one kind of amino acids 
occupies that site in all sequences selected, to $2.996$, if each kind of amino acid is equiprobable,
the latter being the case at large values of the design temperature. 

We have shown elsewhere \cite{isles} that, for monomeric proteins, low--entropy sites are those where most of the 
stabilization energy is concentrated. The search of low--entropy sites rather than of strongly--interacting 
sites has the advantage that it can be easily performed for real proteins, for which a reliable 
energy function is not available. 

In Figs. 11 and 12 we display the entropy per site and the associated distribution of entropy
for: (a) sequences displaying the folding behaviour of sequence $\#1$ (solid squares in Fig. 4), 
(b) for sequences displaying the folding behaviour of sequence $\#3$ (solid triangles in Fig. 4),
(c) aggregating sequences (empty circles in Fig. 4), (d) for the monomer S$_{36}$, single domain
protein, which folds into the native conformation corresponding to half of the homodimer shown in Fig. 3
(sequence $\#6$ of Table I).

One observes a marked difference between the entropy per site shown in Fig. 11(a) and that displayed
in Fig. 11(b).
In fact, sequences of type $\#1$ have many more conserved sites than sequences of type $\#3$ (the ratio
between these two numbers being 3.3), the average entropy being a factor of 1.6 larger than that associated with sequences of type $\#3$. Furthermore, in the case of sequence $\#3$ only $20\%$ of the conserved sites lie
on the surface, while in the case of sequence $\#1$ this ratio is $40\%$.
On the other hand, cases (a) and (b) cannot be distinguished by looking at the interface/bulk average entropy, 
being 1.42 and 1.43 respectively in the case of sequence of type $\#1$ and 1.17 and 1.24 in the case
of sequences of type $\#3$. The situation is similar if one looks at the values of the average entropy
associated with conserved sites lying on the interface ($<S_{cons}>_t$) as compared to the average entropy
$<S_{cons}>$ associated with all conserved sites (cf. Table IV).
 
For aggregating sequences (empty circles in Fig. 4, see Fig. 11(c) ),
there is an evident difference between the entropy of the interface (sites 1--12), whose average entropy 
is 1.18, and the bulk, whose average entropy is 1.71. Aggregation arises due to the fact that the 
surface, which essentially contains all of the "hot" sites of the protein ($85\%$ of them), is too reactive.

Summing up, sequences of type $\#1$ give rise to a dimer where the conservation pattern is distributed 
much more uniformly among all sites and where there are quite a large number of important 
sites although much less conserved than the few hot sites of sequences of type $\#3$.
These results indicate that, from the conservation patterns of lattice designed sequences, it 
is possible to recognize sequences which aggregate from those which dimerize. Among 
these we can recognize sequences which dimerize through a three--state scenario or
through a two--state mechanism by looking at the overall distribution of entropy, but not
at the difference between the entropy of the interface and of the bulk. These results agree with 
the findings by Grishin and Phillips, which analyze the conservation of residues on the 
surface of five oligomeric enzymes and find no signal of any larger conservationism on the 
interface with respect to the bulk \cite{grishin}. We suggest that one has to analyze
the distribution of entropy, not the bulk/interface partition, to derive the behaviour
of a family of sequences.

To test the predictions of the model, we have calculated the entropy at each site for two 
dimeric proteins, i.e. P22 arc repressor and 
aspartate aminotransfease (cf. Figs. \ref{arc} and \ref{asp}). The former folds according to 
the paradigm of sequence $\#1$ of Table I (first dimerize and then fold), while the latter 
follows the opposite scenario (the monomers first fold and then dimerize). From the fssp database we have 
selected, for each of the two proteins, a family of non--homologous aligned sequences which have the 
same fold and the same interface (z--score larger than 2), and from them we have calculated 
$S(i)$. The entropy was calculated in 
three different ways, in order to take care of the gaps in the alignment. The results indicated
in Figs. 1 and 2 in terms of solid lines were 
obtained considering the positions in the gaps as filled by a kind of amino acid which is,
 by definition, different from all others (including the other "fake" amino acids put in the
 gaps of the other sequences). Dots indicate a calculations where amino acids are grouped
into six classes (cf. ref. \cite{leo}), while the  dotted line is calculated ignoring 
sequences which, at a given 
site, display a gap. Below the x--axis it is displayed, with a black stripe, the interface 
region. 

The relative number of conserved sites is considerably larger (a factor 2.7, cf. Table IV)
in the case of P22 arc repressor than in the case of aspartate aminotransfease. Of them, only
$20\%$ lie on the interface of aspartate aminotransfease while $37\%$ does in the case of
the two--state P22 arc repressor.

Further confrontation between the results of lattice model calculations 
and real proteins can be carried out through
the entropy distributions associated with Fig. 11 (designed proteins) and Figs. 1 and 2 
(P22 arc repressor and
aspartate aminotransfease), as displayed in Figs. 12 and 13, respectively.
The arc repressor distribution
displays an abrupt increase of the entropy at a value somewhat larger than 1 which is the 
edge of a well defined peak\footnote{The lowest entropy bin, separated from the others, 
contain a single site and could be associated with functional purposes. The highest bin, on the
other hand, contains all the sites which have gaps in the alignment.}, a behaviour which
is very similar to that of sequences of kind $\#1$ shown in Fig. 12(a). On the other hand, aspartate aminotransfease
displays a gradual increase in the distribution of entropy, and a two peak structure, which
resembles the corresponding quantity associated with sequences of type $\#3$ (cf. Fig. 12(b) ), having
few sites which are highly conserved.
These results, as well as those displayed in Table IV, thus seem to confirm the overall picture emerging from lattice model calculations.

We conclude this section by recalling the fact that the data 
used to derive the numbers displayed in Figs. 1 and 2
(6 and 11 sequences respectively) have rather poor statistics, 
because not many analogous sequences with the same 
interface are available in the PDB. Another caveat to be used in comparing the data with the 
model calculations is the fact that real proteins aside from structural features, 
display functional properties (not present in model calculations), which can have 
conditioned the conservation of amino acids at precise sites, in particular 
surface and interface sites.

\section{Conclusions}

In the present paper we have designed, with the help of a lattice model, sequences which either first dimerize
and then fold or, conversely, first fold and then dimerize. 
They were obtained  by minimizing their energy in a given (dimeric) conformation with respect to amino acid 
sequence at constant composition. The swapping of amino acids was carried out making use of Monte Carlo techniques. 
Two design temperatures have been used to carry out the minimization process, one controlling the 
evolutionary pressure on the interface amino acids, the other controlling the pressure on the amino acids occupying the bulk of the dimer.

We know that the way evolution solves the protein folding problem of small monoglobular systems, 
at least within lattice models, is by asigning the commanding role of the process to few, hot, 
highly conserved amino acids (low entropy "bump" of Fig. 12(d) ). For the remaining amino acids one 
can essentially choose anyone among the twenty different types. In fact, we know \cite{prl} 
that there are $\approx 10^{30}$ 
sequences sharing the same "hot" amino acids which fold to the native structure on 
which the monomeric sequence S$_{36}$ (sequence $\#6$, Table I) was designed (high entropy peak of Fig. 12(d)).

The further requirement of dimerization after folding, typical of sequences of type $\#3$, 
seems to be solved by evolution through an increase in the number of commanding amino acids, 
that is, by increasing the number of highly conserved, low entropy monomers. In other words, 
by shifting amino acid (sites) from the high--S peak of Fig. 12(d) into the low--S peak of the 
same figure, as testified by the entropy distribution associated with sequences of type $\#3$ shown in Fig. 12(b)
(cf. also Figs. 7(b) and 7(d) ).

This change in the strategy of evolution is complete in the case of sequences of type $\#1$, which first dimerize and 
then fold. In this case, all amino acids become essentially equally conserved, the small high--S peak of Fig. 12(b) 
being absorbed in the long tail of the single peak of Fig. 12(a). This is consistent with the low $\varphi$--values observed for the transition state of P22 arc repressor \cite{walburger}.

From these results one can conclude that sequences which first dimerize and then fold must be much more difficult 
to come about than sequences that first fold and then dimerize, in keeping with the fact that the first type 
of sequences are dependent on a significantly larger set of conserved amino acids to reach the protein native structure
in the folding process, than the second type of proteins.

%%%%%%%%%%%%%%%%%%%%%%%%%%%%%%%%%%%%%%%%%%%%%%%%%%%%%%%%%%%%%%%%%%%%%%%%%%%%%%%%%%%%%%%%%%%%%%%%%%%
%% APPENDICES
%%%%%%%%%%%%%%%%%%%%%%%%%%%%%%%%%%%%%%%%%%%%%%%%%%%%%%%%%%%%%%%%%%%%%%%%%%%%%%%%%%%%%%%%%%%%%%%%%%%

\newpage
\centerline{\LARGE{Appendices}}
\appendix

\section{On the Selection of Sequences}

The unusual feature of the present design process is the
fact that it requires different average energies for the bulk and
for the interface. As a consequence, it is not possible to apply
standard equilibrium thermodynamics. 
Anyway, one can still proceed in parallel with the canonical 
ensemble picture and regard the system as composed of two
interacting parts, each of them in contact with its own thermal
bath. Of course we are not interested in the true equilibrium
of the system, which would require the two baths to reach the
same temperature and the same average energy, but in a stationary
state in which the average energies are constant.

If we call $p$ a generic distribution of states of the system, 
we can define the average energy functional $E_1[p]=\sum H_1 p$ and 
$E_2[p]=\sum H_2 p$ of the two parts of the system and the
entropy functional $S[p]=-\sum p\log p$, which indicates the 
information we have about the system.

Among all possible distributions $p$, we are interested in the 
stationary distribution $p^*$ which minimizes the information
(maximizing $S$) at fixed $E_1[p]=E_1^*$ and $E_2[p]=E_2^*$.
In keeping also with the constrain $\sum p=1$, one finds the
distribution
\begin{equation}
p^*=\frac{1}{Z}\exp(-\alpha H_1-\beta H_2),
\end{equation}
where $Z=\sum\exp(-\alpha H_1-\beta H_2)$ normalizes the probability
to unity and $\alpha$ and $\beta$ are the Lagrange multipliers which
set the average energies. To evaluate them, we can write the entropy
$S[p^*]$ as a function of the energies $S(E_1,E_2)=\alpha E_1 +
\beta E_2 - \log Z$. Its derivatives in $E_1^*$ and $E_2^*$ give
\begin{eqnarray}
\frac{\partial S}{\partial E_1}=\alpha; && \frac{\partial S}{\partial E_2}=\beta.
\end{eqnarray}
In parallel with equilibrium thermodynamics, we call temperatures the
inverse of the two Lagrange multipliers, $\tau_1=\alpha^{-1}$ and $\tau_t=\beta^{-1}$. 

To select sequences $\{s\}$ distributed according to 
\begin{equation}
\label{dis_dim}
p(\{s\})=\frac{1}{Z}\exp\left(-\frac{H_1}{\tau_1}-\frac{H_t}{\tau_t}\right), 
\end{equation}
where $H_1$ and $H_t$ are the bulk and the interface energy of the sequence, 
respectively, and
$Z=\sum_{\{s\}}\exp(-H_1/\tau_1-H_t/\tau_t)$, we use a multicanonical technique
(see Appendix B). 

The multicanonical method \cite{multic} is an extension of the usual Monte Carlo sampling 
method. In the latter the phase space of a generic system kept at temperature $T$
is sampled making trial random moves
(here, amino acid swappings) and accepting the move with a
probability $\min(1,\omega(trial)/\omega(old))$, where $\omega(state)$ is the equilibrium 
statistical weight of a given state of the system, which is of course the Boltzman 
distribution $\exp(-E(state)/T)$. This works efficiently at high temperatures,
but becomes problematic when $T$ is decresed, due to the fact that the system
can get trapped in local energy minima. To overcome this problem, the multicanonical
method samples the phase space using unconventional statistical weigths, 
namely $\omega(state)=1/g(state)$, where $g(state)$ is the energetic degeneracy of that state. 
Coarse graining the description of the system, each state can be labelled with its
energy and the statistical weight associated with a given energy is $g(E)\omega(E)=1$. In other
words, we are sampling a phase space which (using these artificial weights) is flat, so that the
system cannot get trapped in metastable states. The problem is that $g(E)$, which defines the
weights, is not known {\it a priori}.
The algorithm has to be, accordingly, self--consistent: a trial $g(E)$ is guessed, 
the phase space is sampled with weights $\omega(E)=1/g(E)$, from the results of the
sampling a new distribution $g(E)$, and so on. When $g(E)$ has converged, one has found the
distribution of energies of the system (which does not depend on the weights used to find it!)
and from this derive all the other thermodynamical quantities of the system.

The algorithm is computationally requiring, also because at each autoconsistent step the
information found in the preceding steps, save the last, is discareded. To solve this 
problem a very efficient method, which combines the informations found during the whole 
calculation, has be developed by Borg and described extensively in ref. \cite{jesper}.
In the sampling of the space of sequences we have used the multicanonical algorithm and
checked the results with the extension of ref. \cite{jesper}.

\section{Simulation of the Dynamics}

Strictly speaking, the Monte Carlo algorithm was 
designed to study equilibrium properties of systems with many degrees of
freedom \cite{metropolis}. Nonetheless, it has been shown
\cite{kikuchi} that, being equivalent to solving the Fokker--Planck
Equation for diffusion in a potential, it can be helpful also 
in studying the kinetical
properties of complex systems, provided that the Fokker--Planck approximation is valid 
(i.e., moves are local and the potential changes smoothly 
on the diffusionlenght scale). Furthermore, Rey and Skolnick
have shown \cite{dynamicmc} that the folding trajectories obtained for single domain
proteins with Monte
Carlo simulations are consistent with those obtained with real Molecular
Dynamics calculations.

In the study of dimers there is an additional problem. Since the Monte Carlo moves are local,
it is not evident that this algorithm describes properly the diffusion of one chain with 
respect to the other.
To make sure that the present algorithm is suitable to deal with diffusion, we have simulated
the displacement of the center of mass of single monomer heteropolymers. In Fig. \ref{com1}
it is displayed the mean square displacement of the center of mass $<x^2>$ as a function
of time for five sequences
designed, to different degrees, to fold to the same conformation. It is also displayed the
behaviour of a random sequence (the first to the left). The calculations are performed at
the temperature $T=0.28$ and the average is done over 50 independent runs, each time starting
from a random conformation. All these sequences move in a diffusive
regime, characterized by $<x^2>=Dt$, where $t$ is time (measured as number of MC steps) and
$D$ is the diffusion coefficient. 

In the inset we show the diffusion coefficient $D$ of each sequence
(in units of $10^{-4}$ lattice units over
number of MC steps) with respect to its native energy (on the horizontal axis is the energy $E_c-E$).
It is clear from this plot that the diffusion coefficient depends on the stability of
the protein, a feature which is rather unphysical since it should only depend on the
shape of the polymer and on the properties of the solvent. On the other hand, one can notice
that for optimized sequences ($E_c-E\geq 0$) $D$ decreases linearly with
the energy, spanning a range $2.11\cdot 10^{-5}MCS^{-1}$ per unit energy. Consequently, we expect
that the effects of this dependence on the folding mechanism of optimized sequences are negligible.

The dependence of the diffusion coefficient on temperature for a fixed sequence (e.g. $\# 1$ of Table I) is
displayed in Fig. \ref{com2}. Except for low temperatures ($T<0.26$) it satisfies Einstein's 
Equation $D=\mu T$, where $\mu$ is the mobility of the chain, indicating that that chain
undergoes Brownian motion.

\section{Periodic Boundary Conditions}

The choice of the linear dimension $L$ of the Wigner cell where the Monte Carlo
simulations were carried out was
found to be important in determining the dynamical evolution of the system. This
quantity is the model parameter which reflects the density $\rho=2/L^3$ of sequences in real systems, 
either the cell or a test tube, which eventually fold in the native conformation of the designed
homodimer. If $L<7$, we found that the chains get entangled in some non--native conformations, since there
is not enough room for them to assume the native conformation (which is a parallelepiped of $6\times 4
\times 3$). If $L>8$ the translational entropy of the disjoint chains is so large that the
conformation corresponding to the two chains folded and separated becomes the equilibrium
state. In other words, although the two chain system is able to reach the native homodimer
conformation, it is very unstable (e.g., 
the relative population of the homodimer native conformation is, for sequence $\#1$ at $T=0.28$ and $L=7$,
less than $10^{-3}$). In what follows we 
shall set\footnote{Due to the discrete character of the conformational space, it is not 
possible to fine--tune the choice of $L$} $L=7$,
even if with this choice, the system experiences some difficulties in reaching the native conformation,
due to the narrowness of the space available. For example,
in 20 long MC runs ($5\cdot 10^8$ MC steps), sequence $\#1$ can find its correct dimeric native state in 16 times out of 20.
In fact, in 4 cases it finds a conformation with energy $E=-35.92$,
corresponding to a situation in which one of the two chains (let us call them A and B)
is folded (say, chain A),
the monomers of chain B at the interface being in their native position, while chain B is in a (well
defined) conformation which has only $40\%$ similarity with its native structure
(similarity parameter $q=0.4$). The reason for this result is to be found in the fact that
chain B builds some contacts with the "back" of chain A, taking advantage of the periodic
boundary conditions. These contacts are mostly between residues of chain B and partners which are
of the right kind but belong to the wrong chain (contacts 17A--32B, 23A--18B, 24A--17B, 25A--36B,
26A--35B, 35A--26B), while two of them are between monomers which cannot be in contact if they
belong to the same chain (31A--32B, 32A--33B). While these processes may be of relevance, for example, in
connection with the formation of amyloid aggregates, it arises in the present work
due to an artifact of the model. In fact, it is connected with
the fact that we are simulating a system of many chains by considering
explicitely only two of them with periodic boundary conditions. As a consequence, a 
domain--swapping--like
mechanism  \cite{aggreg} is likely to take place. In fact, if 
after the two chains have built the native interface, a subdomain of
one of the two sequences is not in place but moves around, 
it can find itself in the vicinity of its complement, but belonging to the other chain.
We shall come back to phenomena of domain--swapping \cite{eisenberg} in a future publication, where we shall consider more
than two chains \cite{dimernew}.

%%%%%%%%%%%%%%%%%%%%%%%%%%%%%%%%%%%%%%%%%%%%%%%%%%%%%%%%%%%%%%%%%%%%%%%%%%%%%%%%%%%%%%%%%%%%%%%%%%%
%% FIGURES
%%%%%%%%%%%%%%%%%%%%%%%%%%%%%%%%%%%%%%%%%%%%%%%%%%%%%%%%%%%%%%%%%%%%%%%%%%%%%%%%%%%%%%%%%%%%%%%%%%%

\newpage

%Fig 1
\begin{figure}
%\centerline{\psfig{file=arc2.eps,height=8cm,width=15cm}}
\caption{The P22 arc repressor (two--state folding [6]) homodimer (right) and the amino acid conservation pattern
for its family of analogous protein (left).}
\label{arc}
\end{figure}

%Fig 2
\begin{figure}
%\centerline{\psfig{file=asp2.eps,height=8cm,width=15cm}}
\caption{The aspartate aminotrasfease (three--state folding [5]) homodimer (right) and the amino acid conservation pattern
for its family of analogous protein (left).}
\label{asp}
\end{figure}

%Fig 3
\begin{figure}
%\centerline{\psfig{file=dimero1_2.eps,height=9cm,width=8cm}}
\caption{Native conformation used to design two (identical) sequences building the homodimer. Each sequence contains 36 amino acids and occupy the vertices of a cubic lattice. To make graphically easier to display the interface contacts (dotted lines) we have artificially (and for the only purpose of design) given these contacts a length which is equal to three times the lattice constant. The native conformation of each of the monomers has been extensively used in the study of the folding of single--domain lattice model proteins \protect\cite{jchemphys1,aggreg,sh_nucleus}}
\label{native}
\end{figure}

%Fig 4
\begin{figure}
%\centerline{\psfig{file=dyn_phases2.eps,height=6cm,width=9cm,angle=-90}}
\caption{The dynamical behaviour (calculated at $T=0.24$) of sequences designed making use of different values of $\tau_1$ (horizontal axis)
and $\tau_t$ (vertical axis). Solid squares indicate sequences that first dimerize and then fold.
Solid triangles indicate sequences that first fold and then dimerize, while empty circles
label sequences which aggregate. The solid line displays the temperatures corresponding
to the energy $E_c$.}
\label{dyn_phases}
\end{figure}

%Fig 5
\begin{figure}
%\centerline{\psfig{file=istoq_a.eps,height=6cm,width=9cm,angle=-90}}
%\centerline{\psfig{file=istoq_b.eps,height=6cm,width=9cm,angle=-90}}
\caption{(a) Distribution of the order parameters $q_1$ (solid curve) and $q_t$ (dashed curve) at
$T=(T^c)_1=0.24$, which is the critical temperature for the formation of the bulk native contacts of sequence $\#1$ of Table I. In the inset, a schematic representation of the behaviour of the average order parameter $<q>$ associated with a first order phase transition through the critical temperature (in the case of an infinite system) is shown. (b) same as above, but for $T=(T^c)_t=0.26$, the critical temperature for the formation of the interface native contacts.}
\label{ph_trans}
\end{figure}

%Fig 6
\begin{figure}
%\centerline{\psfig{file=istoq_2_a.eps,height=6cm,width=9cm,angle=-90}}
%\centerline{\psfig{file=istoq_2_b.eps,height=6cm,width=9cm,angle=-90}}
\caption{Same as Fig. \protect\ref{ph_trans} but for sequence $\#3$ of Table I. (a) Distribution of $q_1$ (solid curve) and of $q_t$ (dashed curve), for $T=(T^c)_t=0.22$, (b) same but for $T=(T^c)_1=0.25$ }
\label{ph_trans2}
\end{figure}

%Fig 7
\begin{figure}
%\centerline{\psfig{file=hot.eps,height=6cm,width=9cm,angle=-90}}
\caption{The average value $\overline{\Delta E_{loc}[m(i)\rightarrow m'(i)]}$. The sites on the surface (1-12) are marked with a heavy line. Also drawn as horizontal dashed lines are the values of $\mu$ and $\mu+2\sigma$ (cf. text).}
\end{figure}

%Fig 8
\begin{figure}
%\centerline{\psfig{file=qt.eps,height=8cm,width=10cm,angle=-90}}
\caption{The distribution of the parameters $q_1$ (continuous curve) and $q_t$ (dashed curve) calculated at the time
in which $q_t$ and $q_1$ reach, for the first time, a value $\geq 0.7$ respectively. The results reported
in (a) correspond to chain $\#1$, while those reported in (b) are associated with chain
$\#3$.}
\label{dynq}
\end{figure}

%Fig 9
\begin{figure}
%\centerline{\psfig{file=unfolding.eps,height=8cm,width=10cm,angle=-90}}
\caption{The distribution of $<q_1(t)>$ (solid curve)
and $<q_t(t)>$ (dashed curve) in the unfolding process, 
starting from the native conformation of the dimer as a function of
t (in units of $10^6$ MC steps) and for $T=0.28$. The continuous and dashed
curve are fits (see text) to the MC results.}
\label{unfold}
\end{figure}

%Fig 10
\newpage
\begin{figure}
%\centerline{\psfig{file=times3.eps,height=8cm,width=10cm,angle=-90}}
\caption{The average folding time $t_f$ as function of the temperature T for
the monomeric sequence $S_{36}$ listed in Table I (sequence $\#6$).}
\end{figure}

%Fig 11
\newpage
\begin{figure}
%\centerline{\psfig{file=entropy_new.eps,height=8cm,width=10cm,angle=-90}}
\caption{The entropy per site calculated for (a) sequence $\#1$, (b) sequence $\#3$,
(c) sequence $\#5$ (aggreg.) and (d) sequence $\#6$ (monomer S$_{36}$). Indicated with a heavy line
the sites belonging to the interface of the dimer.}
\end{figure}

%Fig 12
\begin{figure}
%\centerline{\psfig{file=entropy_isto_new.eps,height=8cm,width=10cm,angle=-90}}
\caption{The distribution $P(S)$ of entropy per site, calculated from Fig. 11.)}
\end{figure}

%Fig 13
\begin{figure}
%\centerline{\psfig{file=isto_arc_asp.eps,height=8cm,width=10cm,angle=-90}}
\caption{The distribution $P(S)$ of entropy per site for the two--state dimer P22 arc repressor (dashed curve),
calculated from Fig. 1, and for the three-state dimer aspartate aminotransfease (solid curve), calculated
from Fig. 2.}
\end{figure}

%Fig 14
\begin{figure}
\caption{The mean square displacement as function of time
(measured as number of MC steps)
of five single domain sequences designed, to different 
extents, to fold to the same three dimensional conformation and 
(the first to the left) of a random sequence. In the inset,
the diffusion coefficient $D=<x^2>/t$ (in units of the square of the lattice
step times $10^{-4}$ divided by the number of MC steps) as a function of the
native energy (the zero of energy is set at $E_c$). The
temperature used in the simulation was $T=0.28$.} 
\label{com1}
\end{figure}

%Fig 15
\begin{figure}
\caption{The diffusion coefficient $D$ as a function
of temperature for the sequence $\#1$ of Table I.}
\label{com2}
\end{figure}

\begin{table}
\caption{In rows 2--5 we display examples of one of the two identical sequences designed, making use of the dimer native conformation displayed in Fig. 3 and of the 20$\times$20 contact energy matrix of ref. [24] (Table VI), at different evolutionary temperatures $\tau_1$ and $\tau_t$. The energy $E_1$ is associated with the bulk contacts of one of the two monomers, so that the total energy in the native conformation is $E_{design}=2E_1+E_t$, $E_t$ being the energy associated with the contacts across the interface. As indicated, sequences $\#1$ and $\#2$ display a two--state dimerization, while sequences $\#3$ and $\#4$ fold through a three-state mechanism. On the other hand, sequence $\#5$ aggregates. Also displayed is the threshold energy $E_c$, as well as the normalized gap $\xi=(E_{design}-E_c)/\sigma$, where $\sigma$ ($=0.3$) is the standared deviation of the contact energies (cf. also footnote number 6). In the last row (number 6) we give, for the sake of comparison, the properties of the monomeric, single domain protein designed on one of the two identical halves of the conformation shown in Fig. 3, and known in the literature as S$_{36}$.}
\begin{tabular}{|c|c|c|c|c|c|c|c|c|c|}
\hline
\# & kind & $\tau_1$ & $\tau_t$ & $E_{1}$ & $E_t$ & $E_{design}$ & $E_c$ & $\xi$ & Sequence \\\hline
1 &{\tiny 2--state}& 0.04 & 0.01 & -15.50 & -5.11 & -36.1 & -35 & 3.7 & {\tiny VLNLGNFVGGHCRYDMEASLWTAKPKPTIRISEADQ} \\
2 &{\tiny 2--state}& 0.01 & 0.01 & -15.29 & -5.04 & -35.6 & -35 & 2.0 & {\tiny NGLVHIFNGGLCRLRMKPSYWTQDAELEAVPDSAKT} \\
3 &{\tiny 3--state}& 0.01 & 0.10 & -16.47 & -3.89 & -36.8 & -35 & 6.0 & {\tiny NTKPVERNCTRVIDGDFALYSGAGSMKLQEHLWPIA} \\
4 &{\tiny 3--state}& 0.10 & 0.10 & -16.54 & -3.91 & -37.0 & -35 & 6.7 & {\tiny NCEVGKFNVLDGSRIHTMAPPLWAQREGLKYADITS} \\
5 &{\tiny aggreg}& 0.10 & 0.01 & -14.95 & -5.11 & -35.0 & -35 & $\approx0$ & {\tiny CGNLVNGHVFGLASMKPRPSDIQWTREAIODYELTA} \\
6 &{\tiny monomer} & 0.01 & & -16.5 & & -16.5 & -14 & 8.3 & {\tiny SQKWLERGATRIADGDLPVNGTYFSCKIMENVHPLA} \\
\hline
\end{tabular}
\label{table1}
\end{table}

\begin{table}
\caption{The folding time $t_f$, the unfolding time $t_1$, the detaching time $t_t$,
and the probability $p_N$ that the protein folds in the case of sequence $\#1$ of Table I. 
In the last row the folding time of the monomer S$_{36}$ is displayed. All times
are expressed in units of $10^6$ MC steps.}
\begin{tabular}{|c|c|c|c|c|}
\hline
T & $t_f$ & $t_1$ & $t_t$ & $p_N$ \\\hline
0.24 & 19  & 0.96 & 0.98 & 0.70\\
0.28 & 24 & 0.79 & 0.58 & 0.45\\
0.32 & 31  & 0.89 & 0.55 & 0.15\\\hline
0.28 & 0.8 & & & 1\\\hline
\end{tabular} 
\end{table}

\begin{table}
\caption{The folding time $t_f$ at different temperatures $T$ for sequence $\#3$ of Table I and
the inverse of the diffusion coefficient (cf. Appendix A) as a function of the temperature.
The times  are expressed in $10^6$ MC steps while while the inverse diffusion coefficients are
expressed in units of time over the lattice unit squared. 
In the last row the folding time of the monomer S$_{36}$ is displayed. }
\begin{tabular}{|c|c|c|}
\hline
T & $t_f$ & $D(T)^{-1}$\\\hline
0.24 & $33$ & $5.8\cdot 10^6$ \\
0.28 & $32$ & $6.6\cdot 10^4$ \\
0.32 & $25$ & $7.1\cdot 10^3$ \\\hline
0.28 & 0.8 & \\\hline
\end{tabular} 
\end{table}

\begin{table}
\caption{The number of conserved sites $N_{cons}$, defined as the number of sites whose entropy
(cf. Fig. 11) fulfills the condition $(S)_{min}\leq S(i)<(S)_{min}+0.25$, divided by the
the total number of residues is displayed in column 4, while the ratio between the number  $N_t$
of conserved sites lying on the surface and the total number of conserved sites $N_{cons}$ is
displayed in column 5. The mean value of the associated entropy is shown in columns 6 and 7. 
In rows 2 and 3 we quote the results associated with three--state dimers, while in the fourth
row we display those corresponding to the monomer S$_{36}$. 
In rows 5 and 6 we quote the results corresponding to two--state dimers,
while in the last row we display the results associated with chain $\#5$ of Table I which aggregate.}
\begin{tabular}{|c|c|c|c|c|c|c|}
\hline
& & $N$ & $(N_{cons}/N)(\%)$ & $(N_t/N_{cons})(\%)$ & $<S_{cons}>$ &  $<S_{cons}>_t$ \\\hline
{\tiny three-state} & aspartate & 600 & $6\%$ & $20\%$ & 1 & $\approx 0.9$ \\\hline
{\tiny three-state}  &$\#3$ & 72 & $8\%$ & $33\%$ & 0.7 &  $\approx 0.7$ \\\hline
 {\tiny monomer} & $\#6$ (S$_{36})$ & 36 &  $8\%$ & & & \\\hline
{\tiny two-state }& arc repressor & 220 &  $16\%$ &  $37\%$ & 1.25 & 1.3 \\\hline
{\tiny two-state} & $\#1$ & 72 &  $28\%$ &  $40\%$  & 1.10 & 1.2  \\\hline
{\tiny aggreg.} & $\#5$ & 72 & $36\%$ & $85 \%$ & 1.1 & 1.2 \\\hline
\end{tabular} 
\end{table}

%%%%%%%%%%%%%%%%%%%%%%%%%%%%%%%%%%%%%%%%%%%%%%%%%%%%%%%%%%%%%%%%%%%%%%%%%%%%%%%%%%%%%%%%%%%%%%%%%%%
%% BIBLIOGRAPHY 
%%%%%%%%%%%%%%%%%%%%%%%%%%%%%%%%%%%%%%%%%%%%%%%%%%%%%%%%%%%%%%%%%%%%%%%%%%%%%%%%%%%%%%%%%%%%%%%%%%%

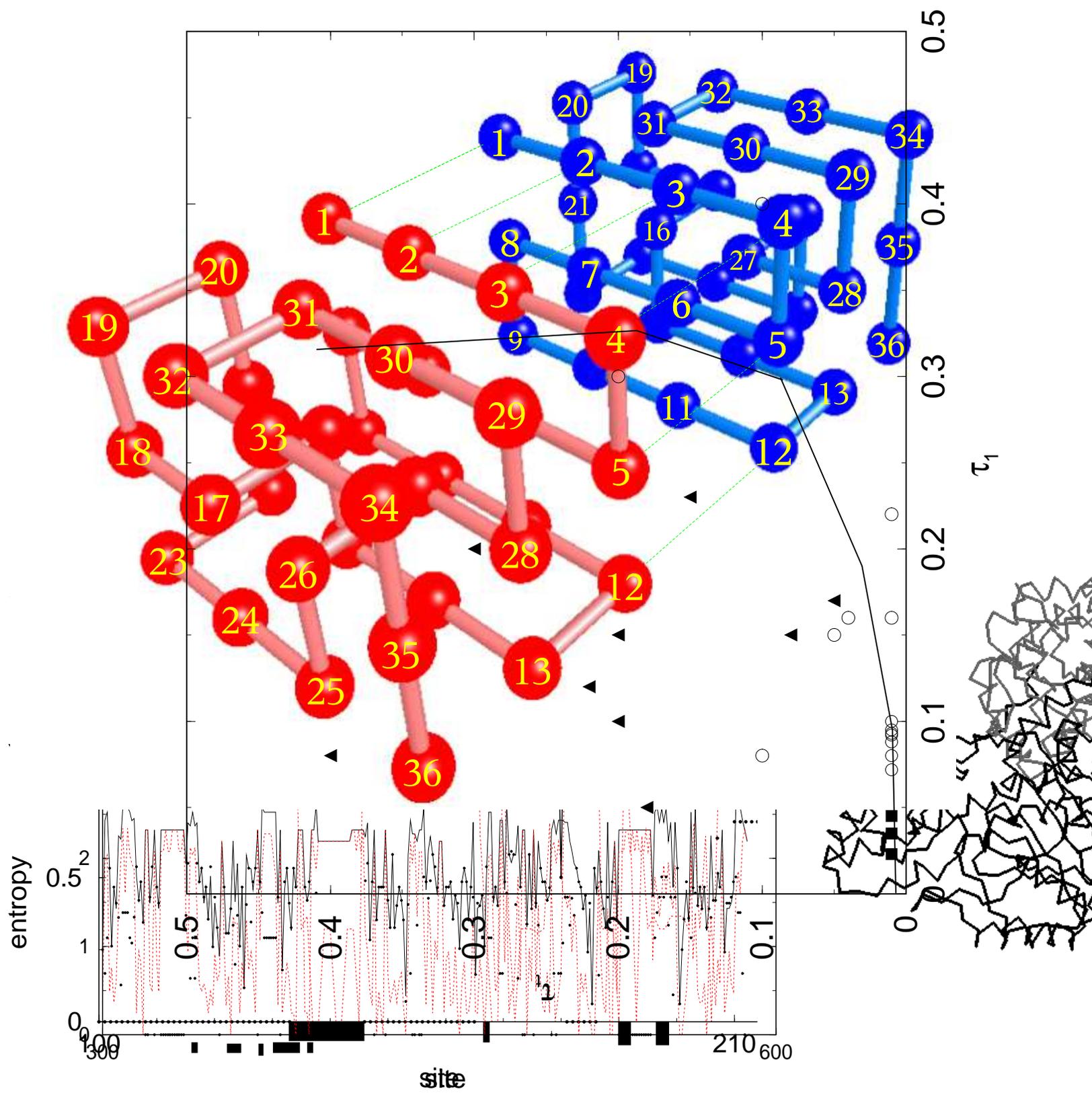
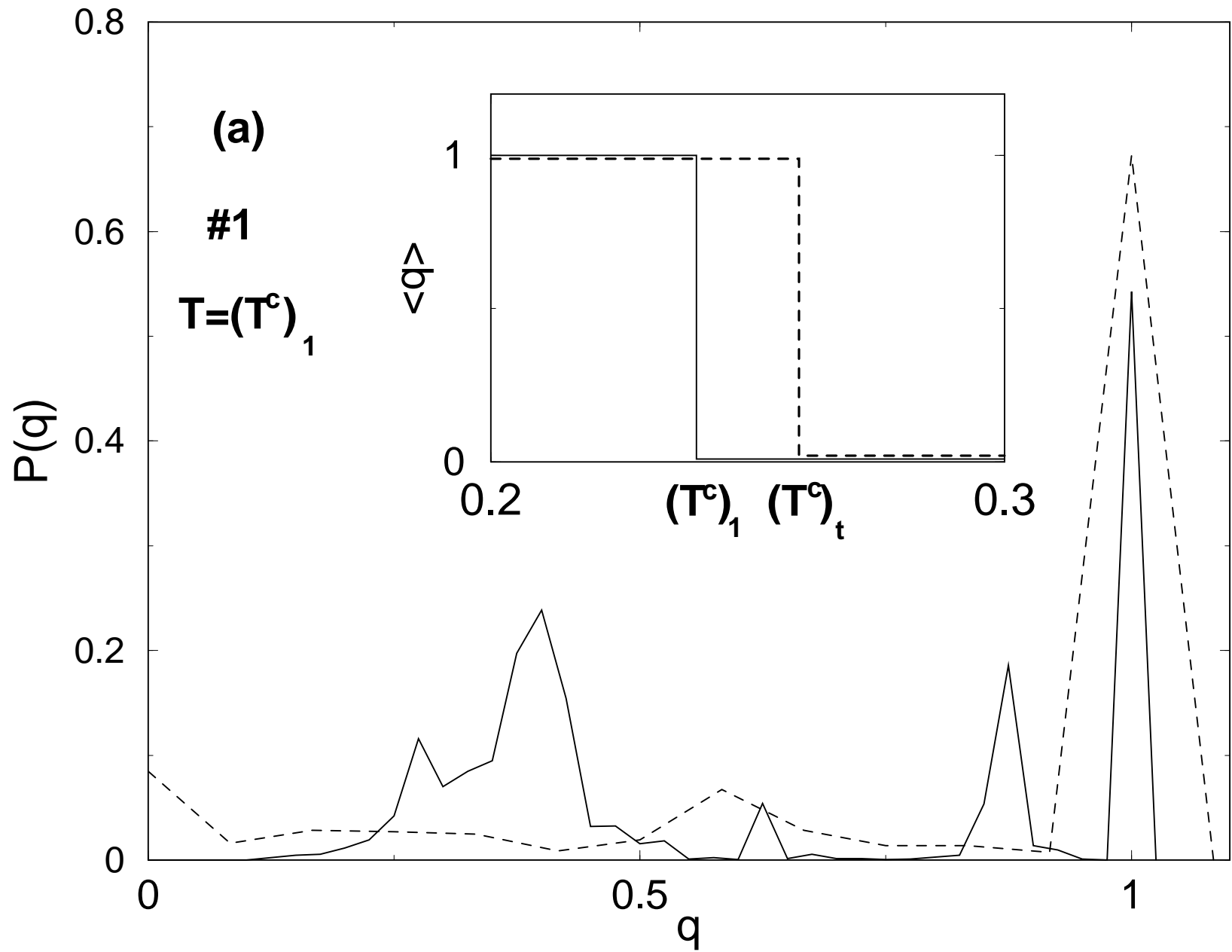
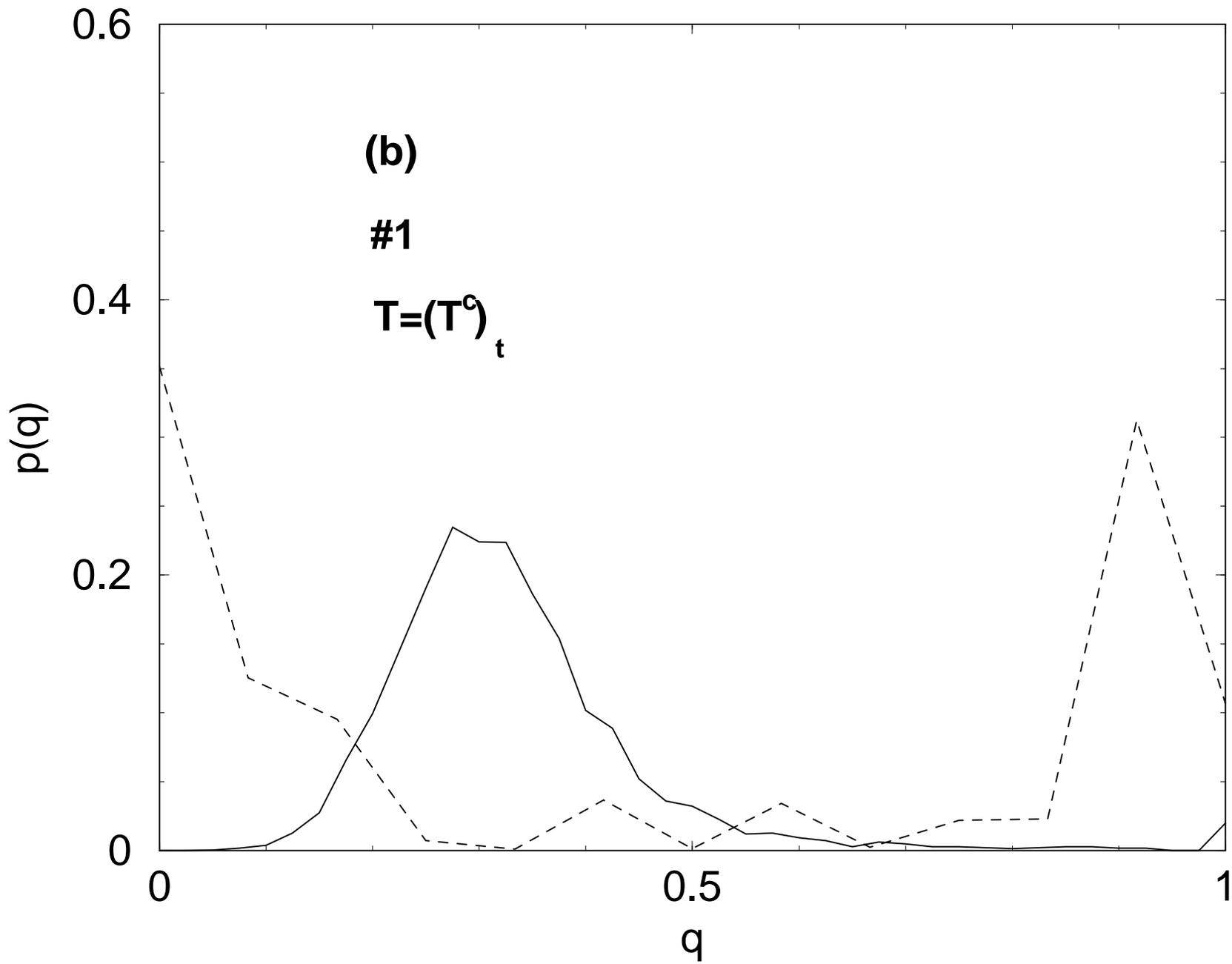
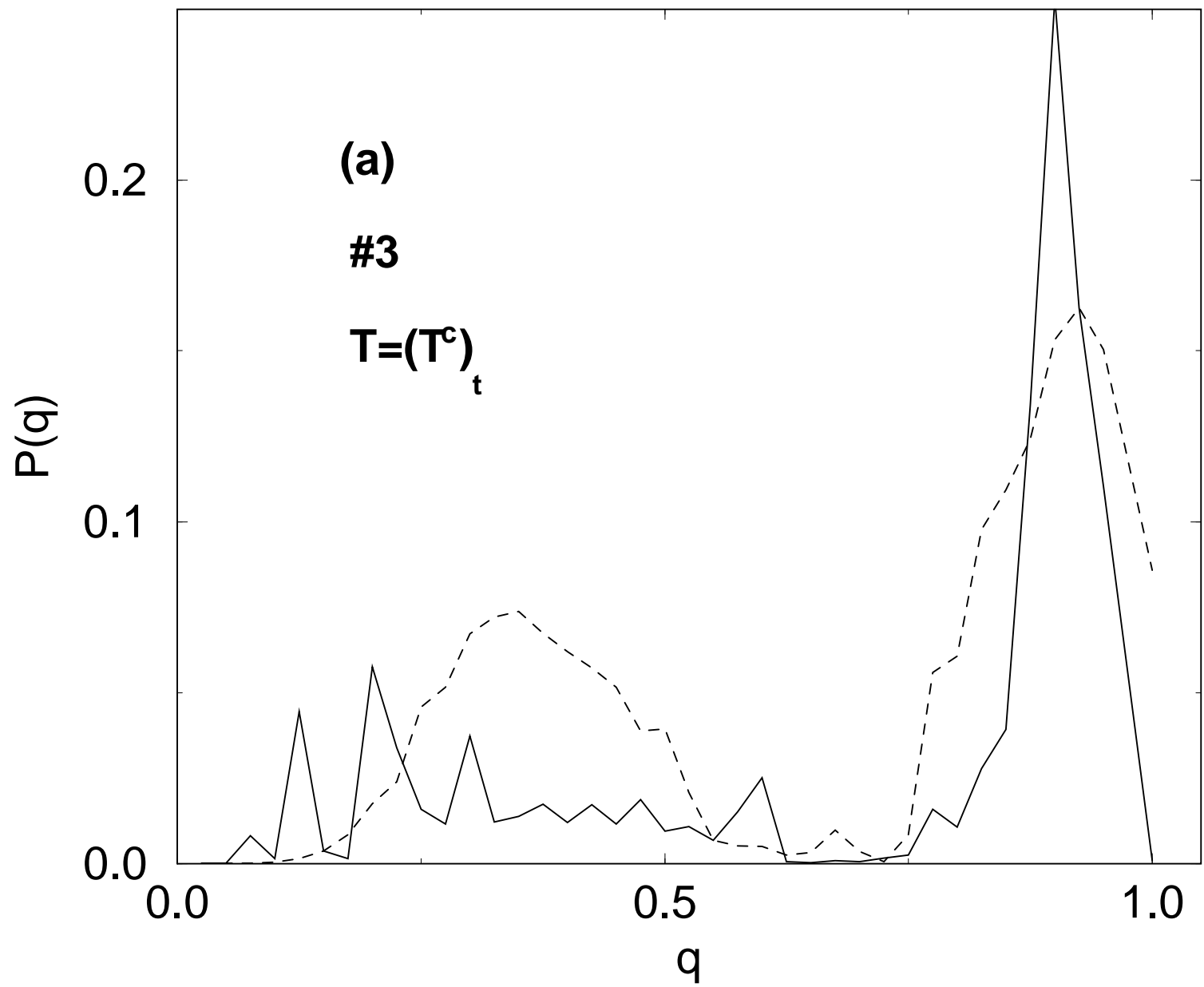
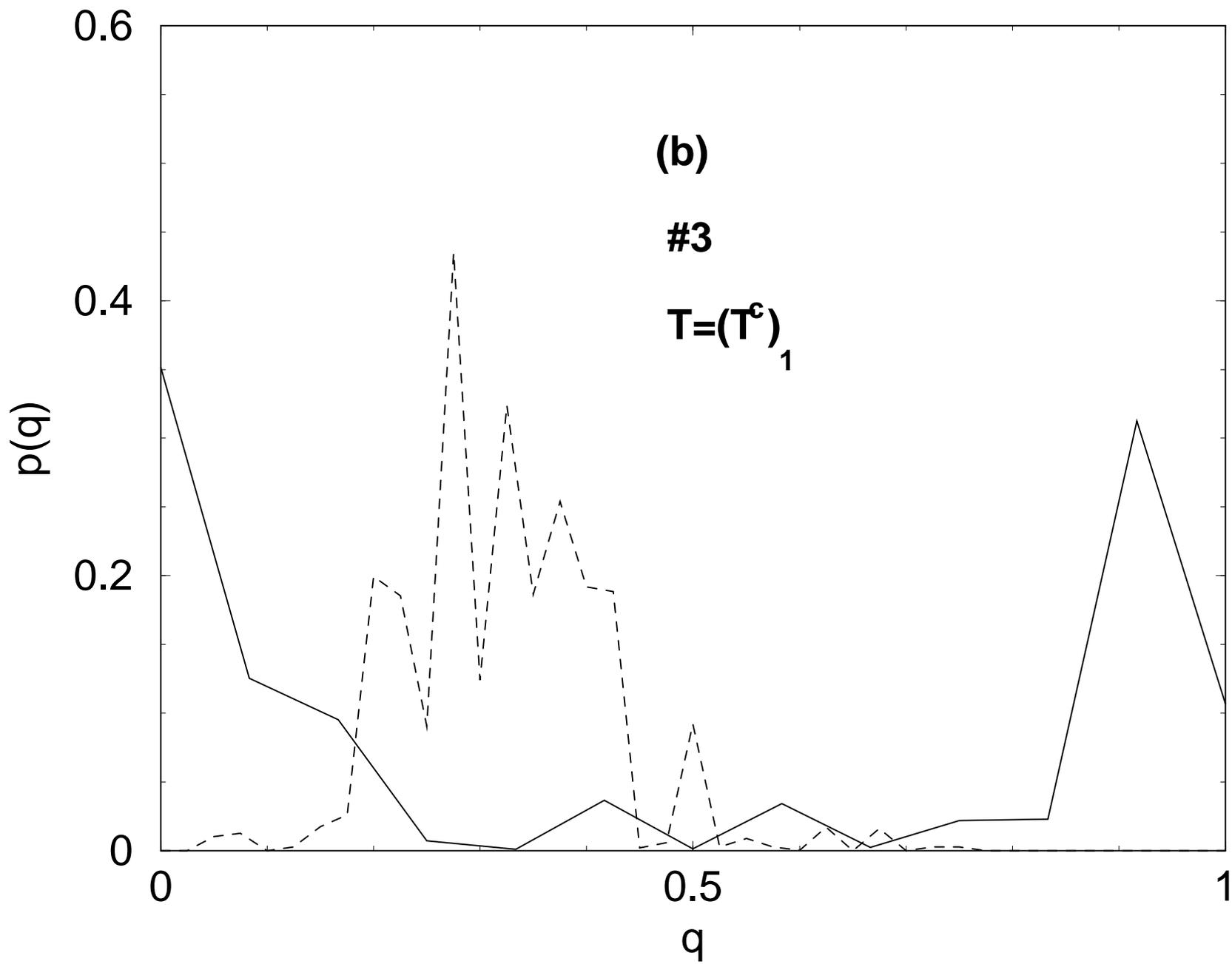
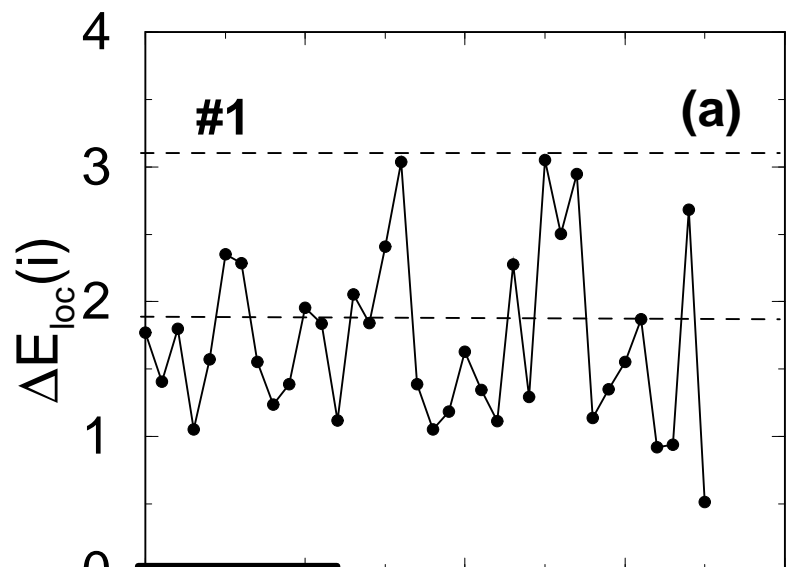
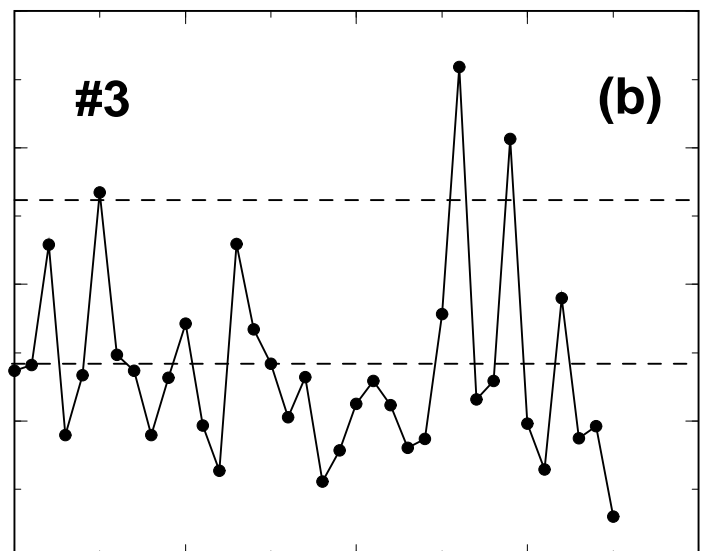
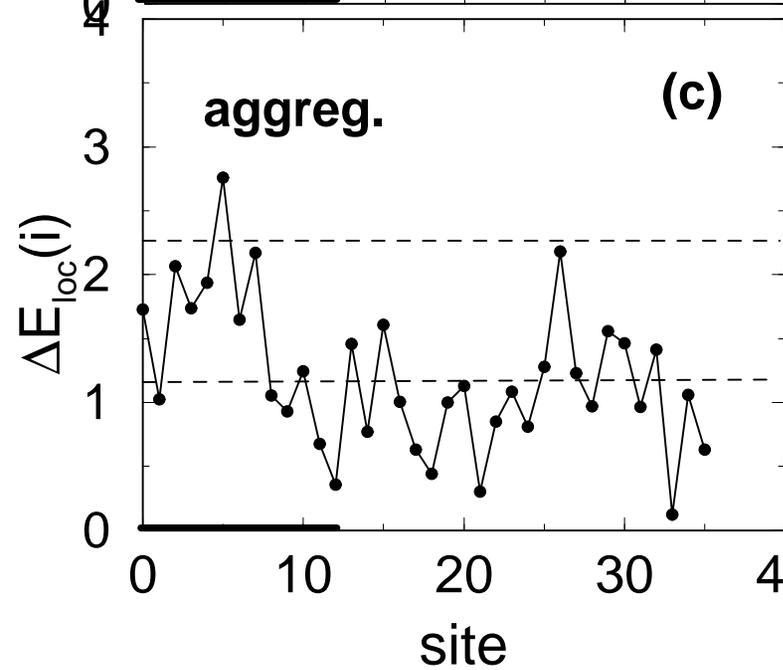
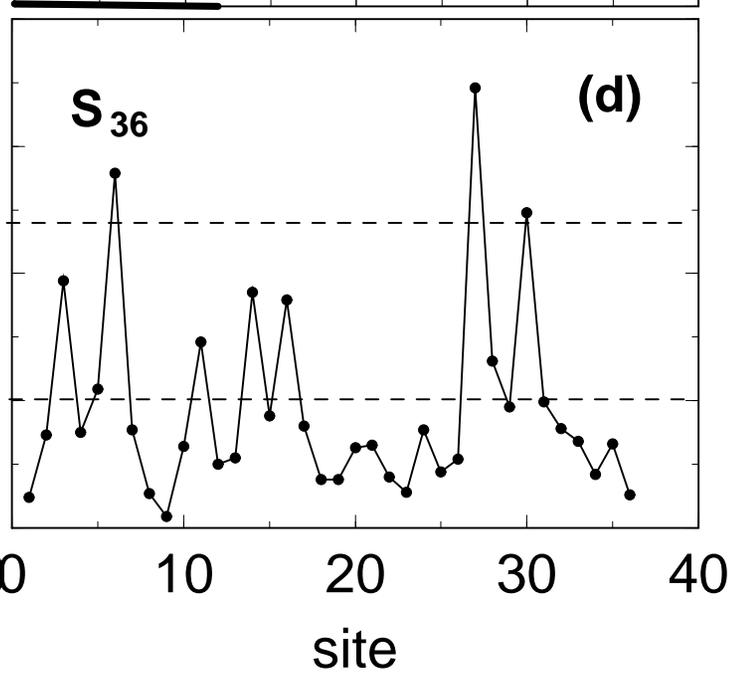
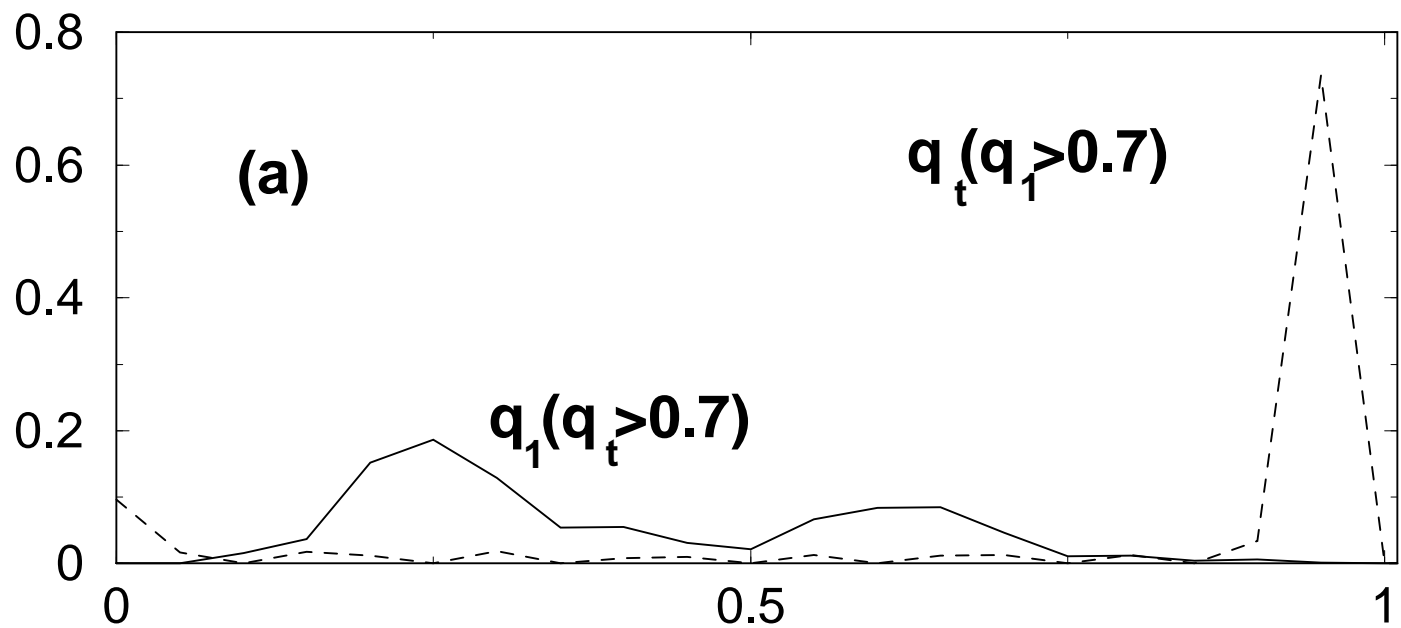
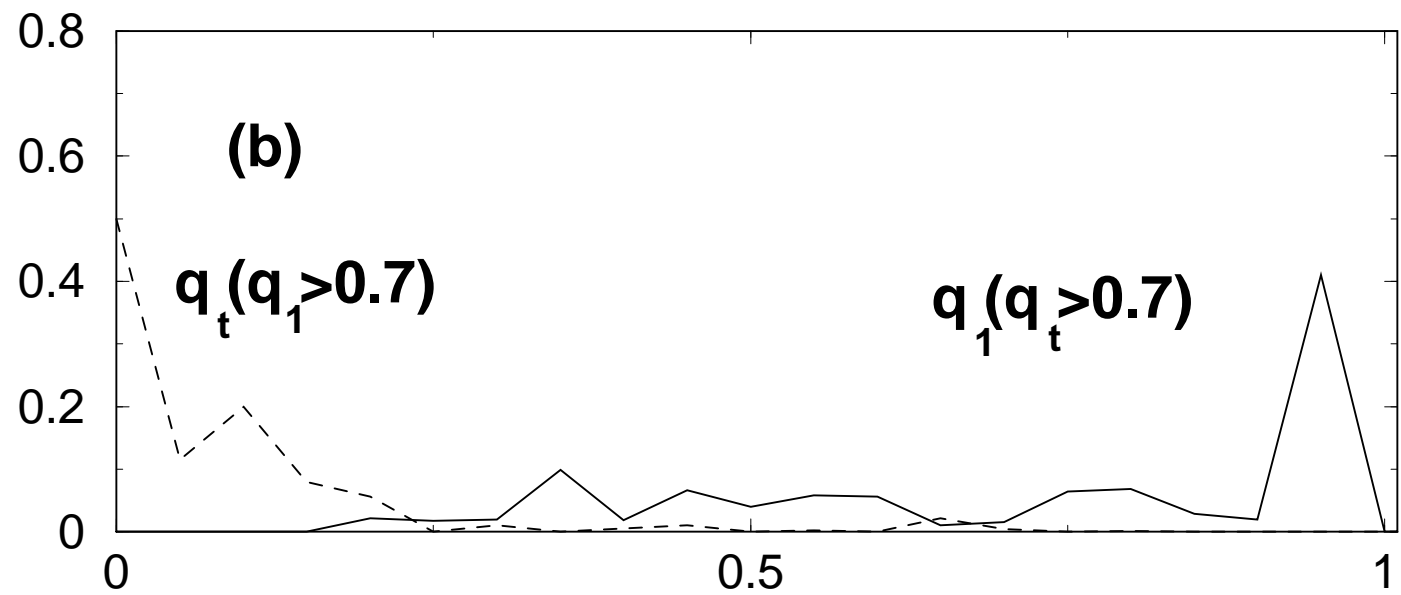
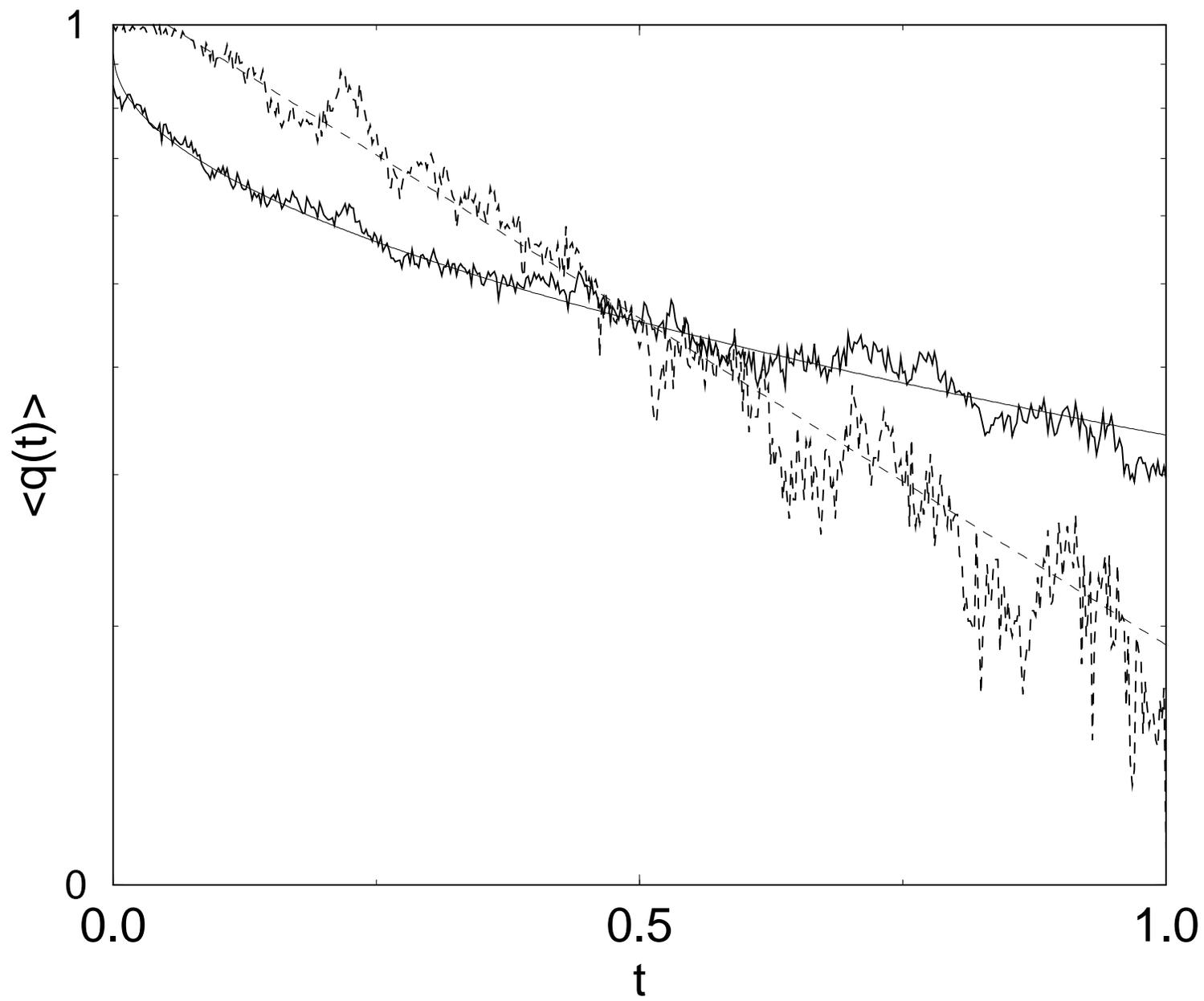
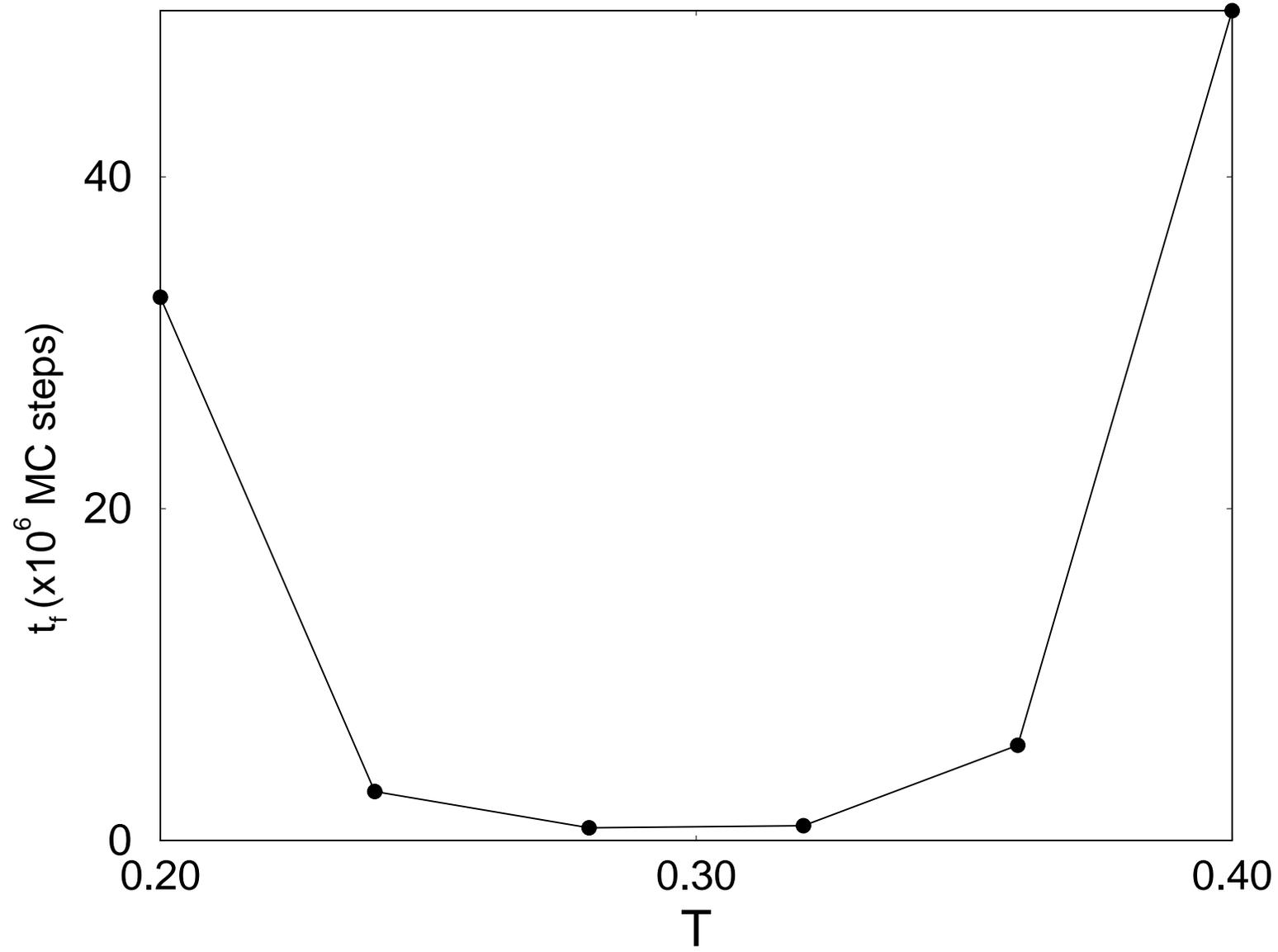
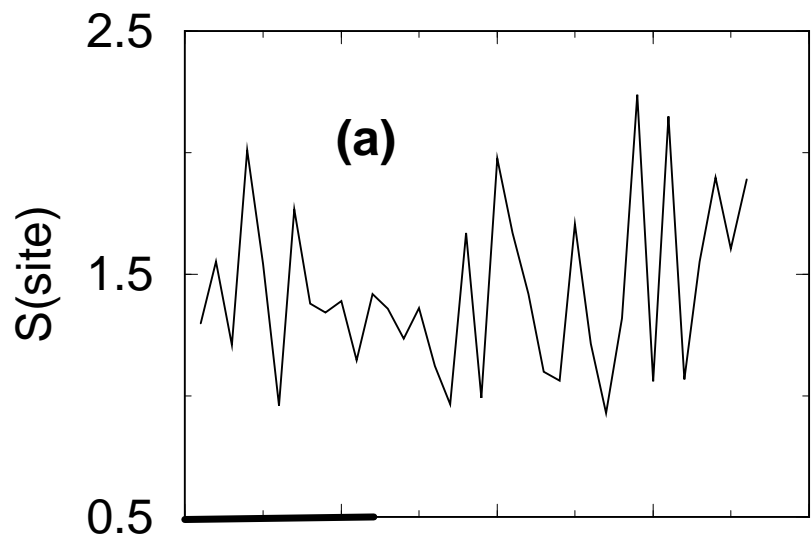
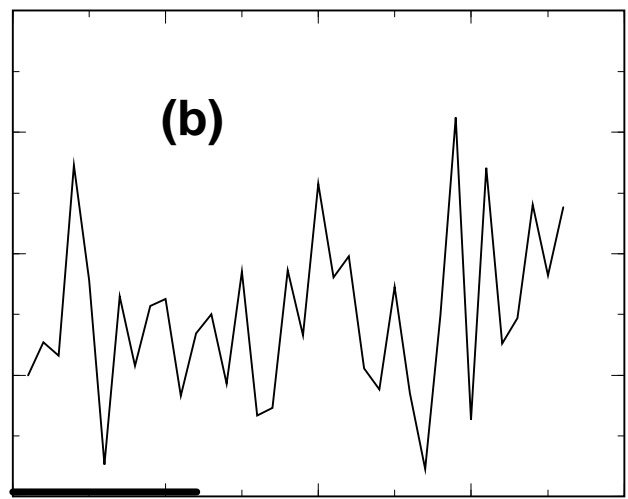
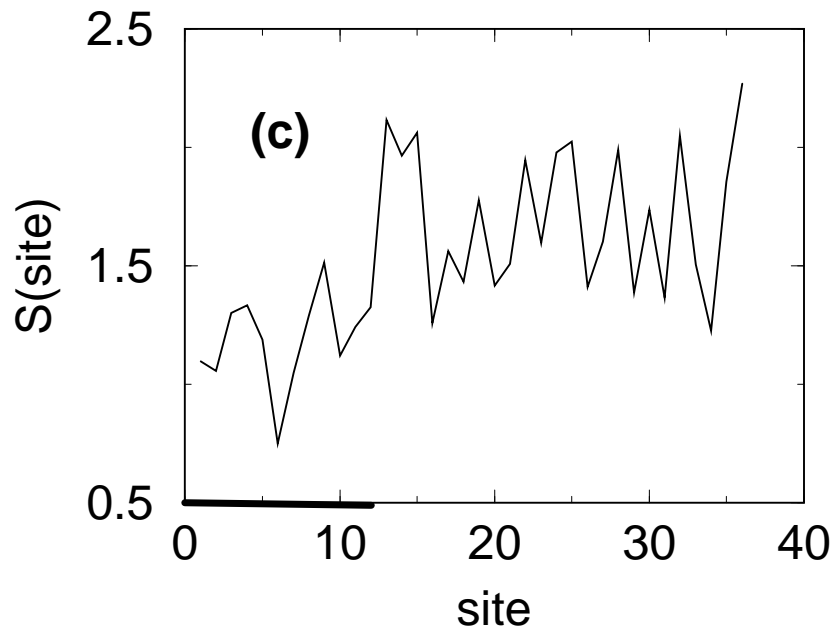
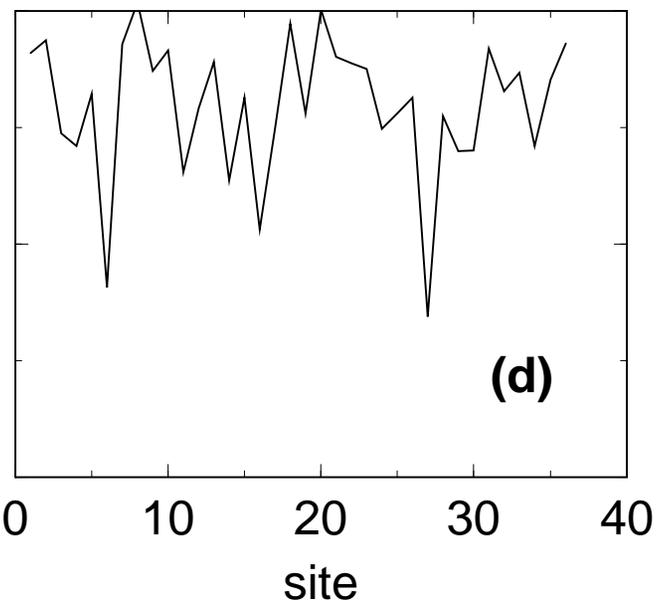
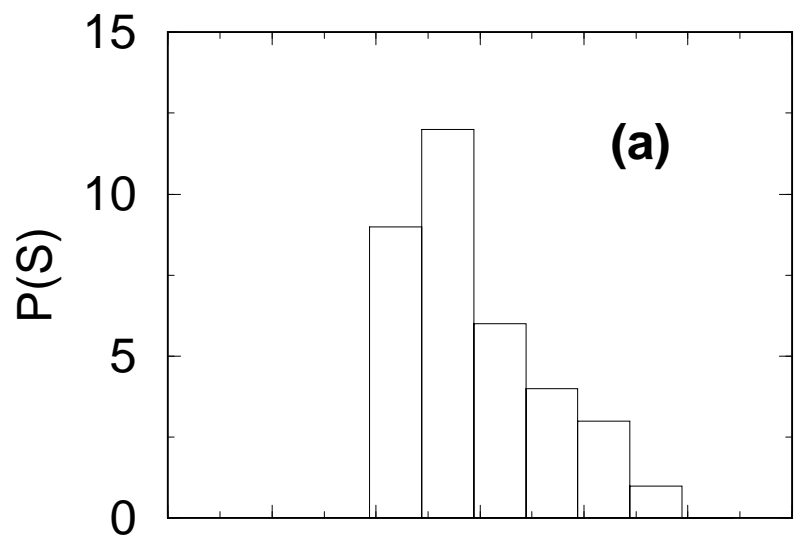
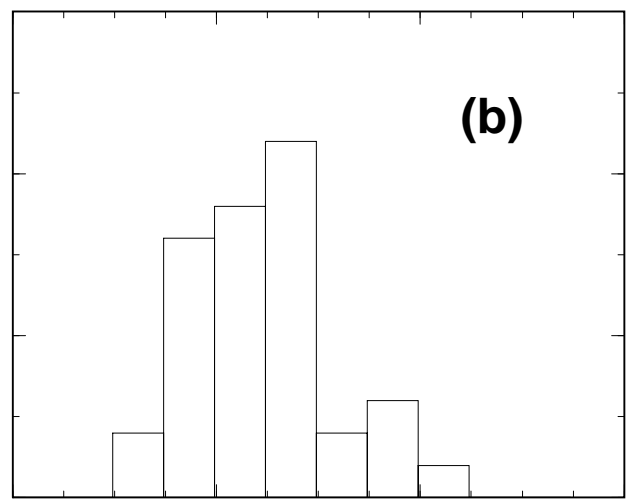
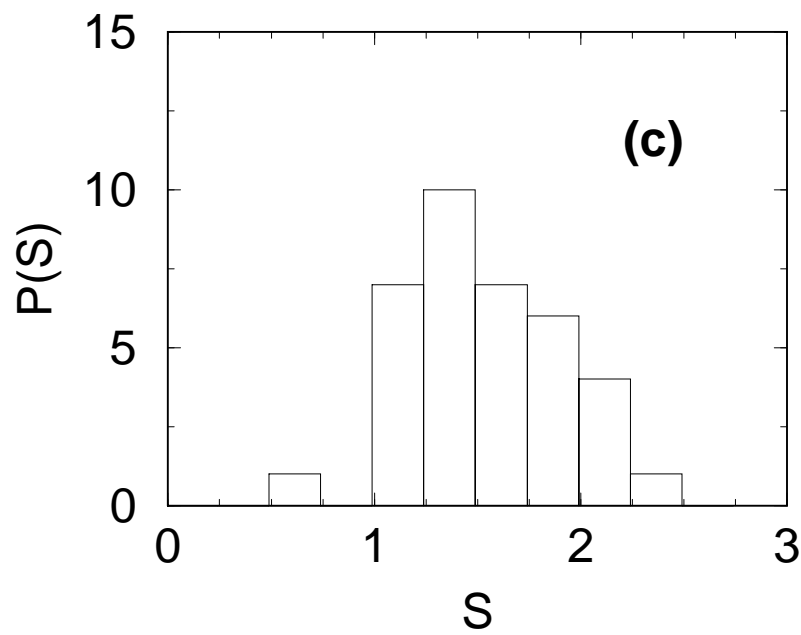
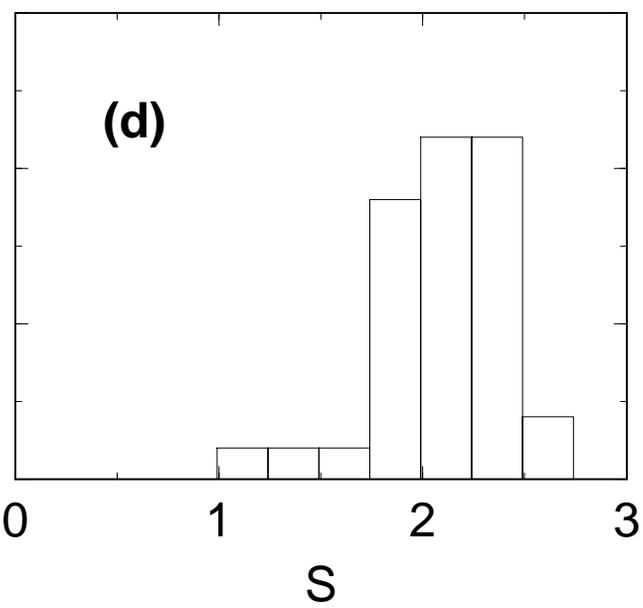
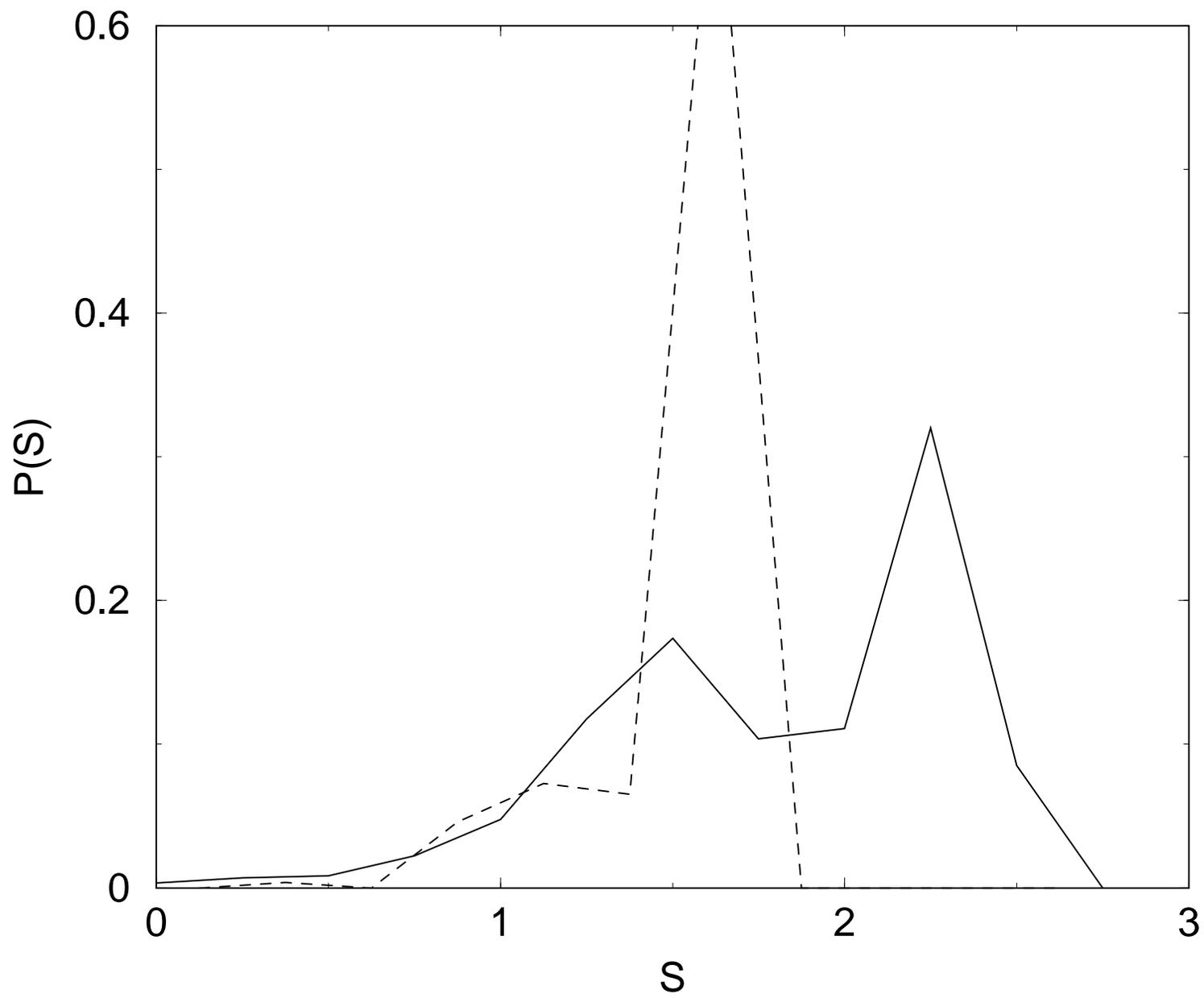
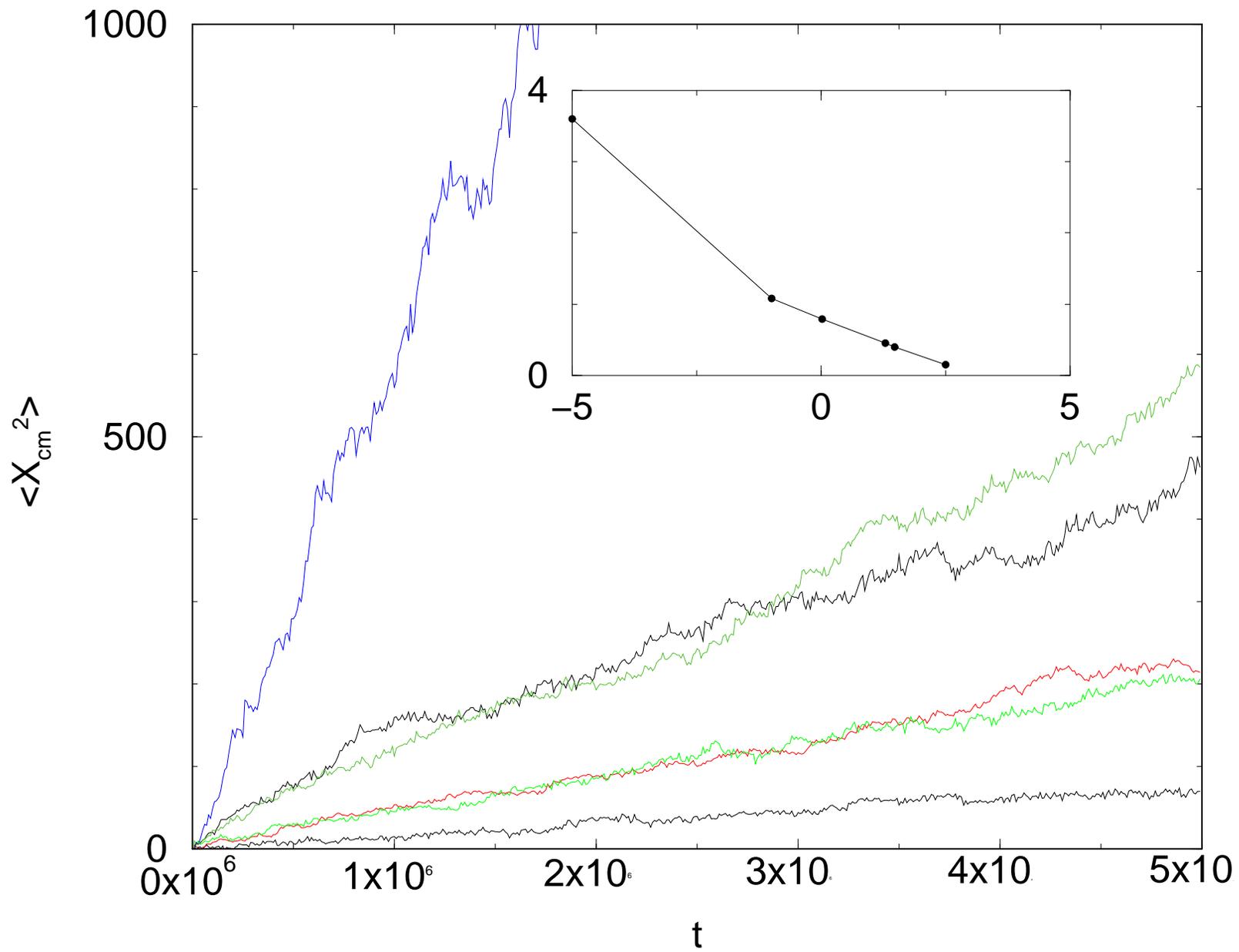
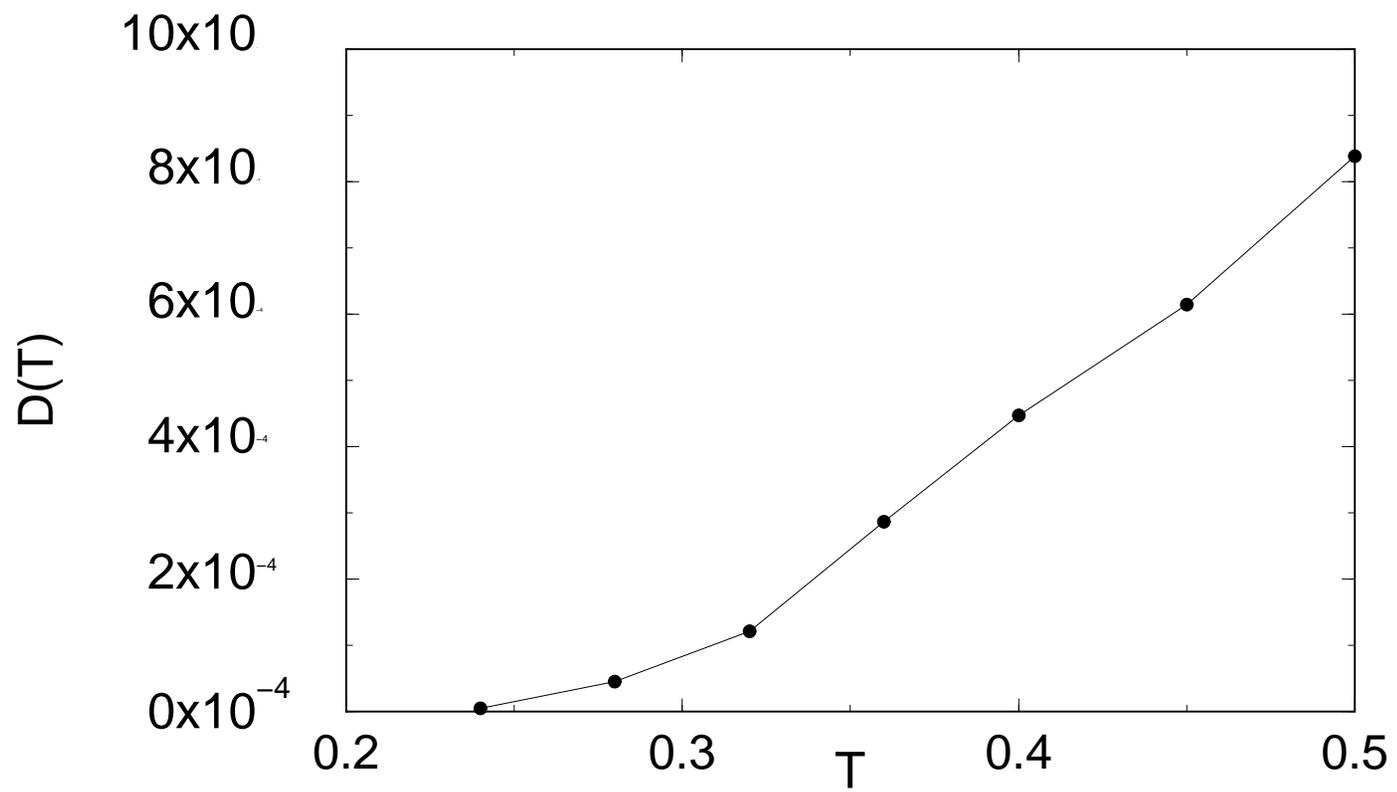
\end{document}